\documentclass[11pt]{article}

\usepackage{cite}
\usepackage{graphicx}
\usepackage{amsmath}
\usepackage{amsmath,amssymb,amsfonts,amsthm,dsfont}
\usepackage{verbatim}
\usepackage{color}
\usepackage{algorithm}
\usepackage{algorithmic}

\allowdisplaybreaks

\theoremstyle{plain} 
\theoremstyle{plain} 
\theoremstyle{plain} \newtheorem{theorem}{\textbf{Theorem}}
\theoremstyle{plain} 
\theoremstyle{plain} \newtheorem{corollary}{\textbf{Corollary}}
\theoremstyle{plain} \newtheorem{definition}{\textbf{Definition}}
\theoremstyle{plain} 
\theoremstyle{plain} 

\newcommand{\isdef}{\stackrel{\textrm{def}}{=}}
\renewcommand{\epsilon}{\varepsilon}
\newcommand{\ket}[1]{| #1 \rangle}
\newcommand{\bra}[1]{\langle #1 |}

\newcommand{\phip}{\ket{\Phi^{+}}}

\newcommand{\squash}[1]{\raisebox{0.04ex}[0pt][0pt]{\small$\textstyle #1$}}
\newcommand{\oosrt}{\squash{\frac{1}{\sqrt{2}}}}

\newcommand\Mark[1]{\textsuperscript#1}

\begin{document}

\title{Exact Classical Simulation \\  of the GHZ Distribution\,
\thanks{A preliminary version of this work has appeared in the \emph{Proceedings of 9th Conference on Theory of Quantum Computation, Communication, and Cryptography} (TQC), Singapore, pp.~7--23, May~2014. Open access at \texttt{http://dx.doi.org/10.4230/LIPIcs.TQC.2014.7}.}}

\author{Gilles Brassard\,\Mark{1}\Mark{,}\Mark{2}, Luc Devroye\,\Mark{3} and Claude Gravel\,\Mark{1}\\[1ex]
\begingroup
\begin{centering}
\begin{tabular}{c}
\small{\Mark{1}\,Universit\'e de Montr\'eal}\\
\small{\Mark{2}\,Canadian Institute for Advanced Research}\\
\small{\Mark{3}\,McGill University}\\[1ex]
\normalsize \texttt{brassard@iro.umontreal.ca},
\texttt{lucdevroye@gmail.com},
\texttt{c03g07@gmail.com}
\end{tabular}
\end{centering}
\endgroup
}

\maketitle
\begin{abstract}
\noindent
John Bell has shown that the correlations entailed by quantum mechanics cannot be reproduced by a classical process involving non-communicating parties. But can they be simulated with the help of bounded communication? This problem has been studied for more than two decades and it is now well understood in the case of bipartite entanglement. However, the issue was still widely open for multipartite entanglement, even for the simplest case, which is the tripartite Greenberger--Horne--Zeilinger (GHZ) state. We give an exact simulation of arbitrary independent von Neumann measurements on general \mbox{$n$-partite} GHZ states. Our protocol requires $O(n^2)$ bits of expected communication between the parties, and $O(n \log n)$ expected time is sufficient to carry it out in parallel.  Furthermore, we need only an expec\-tation of $O(n)$ independent unbiased random bits, with no need for the generation of continuous real random variables nor prior shared random variables.  In~the case of equatorial measurements, we improve on the prior art with a protocol that needs only $O(n \log n)$ bits of communication and $O(\log^2\!n)$ parallel time. At the cost of a slight increase in the number of bits communicated, these tasks can be accomplished with a constant expected number of rounds.\\
\begin{center}
\textbf{Keywords}
\end{center}
Entanglement simulation, Greenberger--Horne--Zeilinger (GHZ) state,
Multipartite entanglement, von Neumann's rejection algorithm, Universal method of inversion,
Knuth-Yao's sampling algorithm.
\end{abstract}

\newpage

\section{Introduction}\label{intro}
\textbf{\Large{T}}he issue of non-locality in quantum physics was raised
in 1935 by Einstein, Podolsky and Rosen
when they introduced the notion of entanglement~\cite{EPR35}.
Thirty years later, Bell proved that the corre\-la\-tions entailed by entanglement
cannot be reproduced by classical local hidden \mbox{variable} \mbox{theories}
between noncommunicating (e.g.~space-like separated) parties~\cite{bell64}.
This~momentous discovery led to the question of \emph{quantifying} quantum non-locality.

A natural quantitative approach to the non-locality inherent in a given
entangled quantum state is
to study the amount of resources that would be required
in a purely classical theory
to \mbox{reproduce} exactly the probabilities corresponding to
measuring it.
More formally, we consider the problem of \emph{sampling}
the joint discrete probability distribution
of the outcomes obtained by people sharing this
quantum state, on which each party applies
locally some measurement on his share.
Each party is given a description of his own measurement
but not informed of the measurements assigned to the other
parties.
This task would be easy (for a theoretician!)\ if the parties were indeed
given their share of the quantum state, but they are~not.
Instead, they must \emph{simulate} the outcome of these measurements
without any quantum resources,
using as little \emph{classical communication} as possible.
\looseness=-1   

This conundrum was introduced by Maudlin~\cite{maudlin92} in 1992
in the simplest case of \mbox{linear} polar\-i\-zation measurements at arbitrary angles on the two photons
that form a Bell state such~as \mbox{$\phip=\oosrt\ket{00}+\oosrt\ket{11}$}\@.
Maudlin claimed that this required ``the capacity~to send messages of~\mbox{unbounded} length'',
but he showed nevertheless that the task could be achieved with a bounded amount of
\emph{expected} communication.
Similar concepts were reinvented independently years later by other researchers~\cite{bct99,steiner00}.
This~led to a series of results, culminating with the protocol of Toner and Bacon
to simulate arbitrary von Neumann \mbox{measurements} on a Bell state with a single bit
of communication \emph{in the worst case}~\cite{tb03}, thus contradicting Maudlin's claim.
Later, Regev and Toner extended this result by giving a
simulation of the correlation (but not the marginals) entailed by arbitrary binary von Neumann measurements
(meaning that the outcome for each party can take only two values)
on arbitrary bipartite states of any dimension using only two bits of communication, also in the worst case~\cite{rt09}.
Inspired by Steiner's work~\cite{steiner00}, Cerf, Gisin and Massar
showed that the effect of an arbitrary pair of positive-operator-valued measurements (POVMs)
on a Bell state can also be simulated with a bounded amount of expected communication~\cite{CGM00}.
A~more detailed early history of the simulation of quantum entanglement
can be found in Ref.~\cite[Sect.~6]{QCC03}.

All this prior work is concerned strictly with the simulation of \emph{bipartite} entanglement.
Much less is known when it comes to simulating multipartite entanglement with classical communication,
a topic that was still teeming with major open problems.
Consider the simplest case, which is the simulation of independent arbitrary von Neumann measurements
on the tripartite GHZ state, named after
Greenberger, Horne and Zeilinger~\cite{GHZ89}, which we shall denote
\mbox{$| \Psi_3 \rangle = \raisebox{0.04ex}[0pt][0pt]{\small$\textstyle \frac{1}{\sqrt{2}}$} | 000 \rangle + \raisebox{0.04ex}[0pt][0pt]{\small$\textstyle \frac{1}{\sqrt{2}}$} | 111 \rangle$},
or more generally on its $n$-partite generalization
\mbox{$| \Psi_n \rangle = \raisebox{0.04ex}[0pt][0pt]{\small$\textstyle \frac{1}{\sqrt{2}}$} | 0^n \rangle + \raisebox{0.04ex}[0pt][0pt]{\small$\textstyle \frac{1}{\sqrt{2}}$} | 1^n \rangle$}.

The easiest situation arises in the special case of \emph{equatorial} measurements
(defined in Section~\ref{sampling})
on the GHZ state because all the marginal probability distributions obtained by
tracing out one or more of the parties are uniform. Hence, it suffices in
this case to simulate the \mbox{$n$-par}\-tite correlation.
Once this has been achieved, all the marginals can easily be made uniform~\cite{gp10}.
Making the best of this observation, Bancal, Branciard and Gisin
have given a protocol to simulate equatorial measurements on the
tripartite and fourpartite GHZ states at an expected cost of
10 and 20~bits of communication, respectively~\cite{BBG10}.
Later on, Branciard and Gisin improved this in the tripartite case with
a protocol using 3 bits of communication in the worst case~\cite{BG11}.
The simulation of equatorial measurements on $| \Psi_n \rangle$ for \mbox{$n\ge 5$}
was handled subsequently by Brassard and Kaplan, with an expected cost of $O(n^2)$ bits
of communication~\cite{BK12}.
This was the best result obtained until now on this line of work.

Despite substantial effort, the case of \emph{arbitrary} von Neumann measurements, even on the
original tripartite GHZ state $| \Psi_3 \rangle$, was still wide open. Here, we solve this problem in
the general case of the simulation of the $n$-partite GHZ state~$| \Psi_n \rangle$, for any $n$,
under the \emph{random bit model} introduced in 1976 by Knuth and Yao~\cite{knuthyao76},
in which the only source of randomness comes
from the availability of inde\-pendently distributed unbiased random bits.
Furthermore, we have no needs for prior shared random variables
between the parties.\,
\footnote{\,Most~of the prior art on the simulation of entanglement by classical communication
required the parties to share \emph{continuous} real random variables
in an initialization phase~\cite{maudlin92,bct99,steiner00,tb03,rt09,CGM00}, admittedly an unreasonable proposition,
but there have been exceptions, such as Ref.~\cite{MBCC01}.}
Our simulation proceeds with $O(n)$ expected perfect random bits and its
expected communication cost is $O(n^2)$ bits, but only
$O(n \log n)$ \emph{time} if we count one step for sending bits in parallel
according to a realistic scenario in which no party has to send or receive more than one bit in any given step.
Furthermore, in the case of equatorial measurements,
we improve the earlier best \mbox{result}~\cite{BK12}
with an expected communication cost of only~$O(n \log n)$ bits and \mbox{$O(\log^2 \!n)$} parallel time.
At~the cost of a slight increase in the number of bits communicated and the number of required random bits,
these tasks can be accomplished with a constant expected number of rounds.

More formally, the quantum task that we want to simulate is as follows.
Each party~$j$ holds one qubit (quantum bit) from
state \mbox{$| \Psi_n \rangle = \raisebox{0.04ex}[0pt][0pt]{\small$\textstyle \frac{1}{\sqrt{2}}$} | 0^n \rangle + \raisebox{0.04ex}[0pt][0pt]{\small$\textstyle \frac{1}{\sqrt{2}}$} | 1^n \rangle$}
and is given the description of a von Neumann measure\-ment~$M_j$. By~local operations, they collectively perform
\mbox{$\otimes_{j=1}^n M_j$} on~$| \Psi_n \rangle$, thus obtaining one outcome each, say~\mbox{$b_j \in \{-1,+1\}$},
which is their output.
The~joint probability distribution $p(b)$ of the~$b_j$'s is defined by the joint set of measurements.
Our~purpose is to sample \emph{exactly} this joint probability distribution by a purely classical process
that involves no prior shared random variables and as little communication as possible.
As~mentioned above, previous solutions~\cite{BBG10,BG11,BK12} required
each individual measurement to be equatorial.
In~order to overcome this limitation, our complete \mbox{solution} builds on four ingredients:
(1)~Gravel's decomposition of $p(b)$ as
a convex combination of two sub-distributions~\cite{gra2011mem,gra2012};
(2)~Knuth and Yao's algo\-rithm~\cite{knuthyao76}
to sample exactly discrete probability distributions assuming only a source of unbiased
identically independently distributed \emph{bits}, rather than a source of \emph{continuous}
uniform random variables on the interval \mbox{$[0,1]$};
(3)~the \emph{universal method of inversion} \cite[for instance]{devbook86}; and
(4)~our own distributed version of the classic \emph{von Neumann's rejection algorithm}~\cite{vN51}.

We define precisely our problem in Section~\ref{sampling} and we formulate our convex decomposition
of the GHZ distribution, which is the key to its simulation. Then, we explain how to sample according
to a Bernoulli distribution even when only approximations of the distribution's parameter are available.
We~also explain how the classic von Neumann rejection algorithm can be used to sample in the
sub-distributions defined by our convex decomposition. However, little attention is paid in
Section~\ref{sampling} to the fact that the
various parameters that define the joint distribution are not available in a single place.
Section~\ref{comcomp} is concerned with the communication complexity issues.
This paves the way to Section~\ref{final-protocol}, in which we provide
a complete protocol to solve our problem, as well as its detailed analysis.
Section~\ref{variations} discusses variations on the theme,
in which we consider a parallel model of communication,
an expected bounded-round solution, improvements on the prior art for the simulation of
equatorial measurements, and a remark to the effect that only one party needs access
to a source of randomness. We~conclude in Section~\ref{discussion} with a discussion, open problems,
and the announcement of a forthcoming generalization of our results to all multiparty entangled states
in which each party is given a single qubit.
For~completeness, the appendices derive from first principles
our convex decomposition of the GHZ distribution,
as well as elementary approximation and truncation formulas useful in the analysis
of the parallel model.

\section{Sampling exactly the GHZ distribution in the random bit model}\label{sampling}
Any von Neumann measurement
on a single qubit can be conveniently represented by a point on the
surface of a three-dimensional sphere, known as the
Bloch sphere, whose spherical coor\-di\-nates can be
specified by an \emph{azimuthal} angle \mbox{$\theta \in [0,2\pi)$} and
an \emph{elevation} angle \mbox{$\varphi \in [-\pi/2,\pi/2]$}.
These parameters define operator
\begin{displaymath}
M = x \, \sigma_1 + y \, \sigma_2 + z \, \sigma_3 =
\left( \begin{array}{cc}
        \sin \varphi  & e^{-\imath\theta} \, \cos \varphi \\
        e^{\imath\theta} \, \cos \varphi  & -\sin \varphi
\end{array}\right) \, ,
\end{displaymath}
where
\mbox{$x = \cos \theta \cos \varphi $},
\mbox{$y = \sin \theta \cos \varphi $},
\mbox{$z = \sin \varphi  $},
and $\sigma_1$, $\sigma_2$ and $\sigma_3$ are the Pauli operators.
In~turn, this operator defines a measurement in the usual way,
which we shall also call $M$ for convenience, whose outcome is one of its
eigenvalues $+1$ or~$-1$.
The~azimuthal angle~$\theta$ represents the equatorial part of the measurement
and the elevation angle $\varphi$ represents its real part.
A~von Neumann measurement is said to be \emph{equatorial} when
its elevation angle \mbox{$\varphi=0$} vanishes and
it is said to be \emph{in the computational basis} when \mbox{$\varphi=\pm\pi/2$}.

Consider a set of $n$ von Neumann single-qubit measurements $M_j$,
represented by their parameters \mbox{$(\theta_j,\varphi_j)$}, \mbox{$1 \le j \le n$}.
This set of operators defines a joint measurement  \mbox{$M=\otimes_{j=1}^n M_j$}.
In~turn, this measurement defines a probability distribution~$p$, which we shall call
the \emph{GHZ distribution}, on the set $\{ -1, +1 \}^n$.
This distribution corresponds to the probability of all possible outcomes when
the $n$-partite GHZ state \mbox{$| \Psi_n \rangle = \raisebox{0.04ex}[0pt][0pt]{\small$\textstyle \frac{1}{\sqrt{2}}$} | 0^n \rangle + \raisebox{0.04ex}[0pt][0pt]{\small$\textstyle \frac{1}{\sqrt{2}}$} | 1^n \rangle$}
is measured according to~$M$.

Following Refs.~\cite{gra2011mem,gra2012}, we show in Appendix~\ref{App-decomp}
that the probability $p(b)$ of obtaining \mbox{$b = (b_1, \ldots, b_n)$} in \mbox{$\{ -1,+1 \}^n$}
can be decomposed as
\begin{equation}\label{eq:p}\textstyle
p(b)=
\cos^2\!\big(\frac{\theta}{2}\big) \, p_1(b)+\sin^2\!\big(\frac{\theta}{2}\big) \, p_2(b) \, ,
\mbox{~where~} \mbox{$\theta = \sum_{j=1}^{n} \theta_j$} \mbox{~and}
\end{equation}
\vspace{-3ex} 
\begin{align}
p_1(b)&=\frac{1}{2}\big(\mathrm{a}_1(b)+\mathrm{a}_2(b)\big)^2,&p_2(b)&=\frac{1}{2}\big(\mathrm{a}_1(b)-\mathrm{a}_2(b)\big)^2,\label{eq:p1p2}\\
\mathrm{a}_1(b)&=\prod_{j=1}^{n} \textstyle {\cos\big(\frac{1}{2}\big(\varphi_j-\frac{\pi}{2}b_j\big)\big)},&
\mathrm{a}_2(b)&=\prod_{j=1}^{n} \textstyle {-\sin\big(\frac{1}{2}\big(\varphi_j-\frac{\pi}{2}b_j\big)\big)}\,.\label{eq:a1a2}
\end{align}
Hence, we see that distribution $p$ is a convex combination of sub-distributions $p_1$ and~$p_2$,
in which the coefficients $\cos^2(\theta/2)$ and $\sin^2(\theta/2)$
depend only on the equatorial part of the measurements,
whereas the sub-distributions depend only on their real part.
Further\-more, it is easy to see that the squares of $\mathrm{a}_1$ and $\mathrm{a}_2$
are themselves discrete probability distributions.

Sampling $p$ is therefore a matter of sampling a Bernoulli distribution with defining parameter
$\cos^2(\theta/2)$ before sampling either $p_1$ or $p_2$, whichever is the case.
Notice that sampling $p_2$ reduces to sampling $p_1$ if, say, we replace
$\varphi_1$ by $\varphi_1 + 2\pi$. As we shall see, full knowledge of the parameters
is not required to sample $p$ \emph{exactly}. We~shall see in Section~\ref{Bernoulli}
how to sample a Bernoulli distribution with
an arbitrary \mbox{$p\in[0,1]$} as parameter (not the same~$p$ as our probability distribution for~GHZ)
using a sequence of approximants converging to~$p$ and using an expected number
of only five unbiased identically independently distributed (i.i.d.)\ random bits.
Subsequently, we shall see in Section~\ref{sampling_p} how to sample $p_1$
by modifying von Neumann's rejection algorithm in a way that it uses sequences
of approximants and unbiased i.i.d.\ random bits. For simulating exactly the GHZ
distribution, an expected number of $6n+17$ perfect random bits is sufficient.

\subsection{Sampling a Bernoulli distribution}\label{Bernoulli}
Assume that only a random bit generator is available to sample a given probability distribution
and that the parameters that specify this distribution are only accessible as follows: we can ask
for any number of bits of each parameter, but will be charged one unit of cost per bit that is revealed.
We~shall also be charged for each random bit requested from the generator

To warm up to this conundrum, consider the problem of generating
a Bernoulli random variable $Y$ with parameter \mbox{$p \in [0,1]$}.
If~$U=0.U_1 U_2\ldots$ is the binary expansion of a uniform $[0,1]$ random variable,
i.e.~$U_1, U_2, \ldots$ is our source of unbiased independent random
bits, and if $p= 0.p_1 p_2 \ldots$ is the binary expansion of~$p$
(in~case \mbox{$p=1$}, we can proceed as if it were $0.p_1 p_2 \ldots$
with each \mbox{$p_i=1$}, and similarly for the probability~$0$ event that \mbox{$U=1$}),
we~compare bits $U_i$ and $p_i$ for $i=1,2,\ldots$ until
for the first time $U_i \not= p_i$. Then, if $U_i = 0 < p_i = 1$,
we return $Y=1$, and if $U_i = 1 > p_i = 0$, we return $Y=0$.
If~we disregard the case $U=p$, which would result in an infinite loop but occurs with probability~0,
it is clear that $Y=1$ if and only if $U < p$.
Therefore, $Y$ is indeed Bernoulli$(p)$ since $U < p$ happens with probability~$p$.
The expected number of bits required from $p$ is precisely~2.
The expected number of bits needed from our random bit source is also~2.

Now, suppose that the parameter $p$ defining our Bernoulli distribution is given by \mbox{$p = \cos^2(\theta/2)$},
as in the case of our decomposition of the GHZ distribution.
None of the parties can know $\theta$ precisely since it is distributed as a sum of $\theta_j$'s,
each of which is known only by one individual party.
If~we could obtain as many physical bits of~$p$ as needed (\mbox{although} the expected number of required bits
is as little as~2), we would use the idea given above
in order to sample
according to this Bernoulli distribution.
However, it is not possible in general to know even the first bit of~$p$
given any fixed number of bits of the~$\theta_j$'s.
(For~instance, if $\theta$ is arbitrarily close to $\pi/2$, we need arbitrarily many bits of precision
about it
before we can tell if the first bit in the binary expansion of $\cos^2(\theta/2)$ is $0$ or~$1$).
Nevertheless, we can
use \emph{approximations} of~$p$, rather than \emph{truncations}, which in turn can come from approximations
(in~particular truncations) of~the~$\theta_j$'s.

\begin{definition}\label{def:approxtrunc}
A~\mbox{$k$-bit} \emph{approximation} of a real number $x$ is any $\hat{x}$ such that $|x-\hat{x}|\leq 2^{-k}$.
A~special case of \mbox{$k$-bit} approximation is the \mbox{$k$-bit} \emph{truncation}
\mbox{$\hat{x} = \mathrm{sign}(x) \lfloor |x| 2^k \rfloor / 2^k$},
where $\mathrm{sign}(x)$ is equal to $+1$, $0$ or $-1$ depending on the sign of~$x$.
Note that the value of $k$ corresponds to the number of bits in the fractional part, without limitation
on the size of the integer part, and that it does not take account of the sign in case it has to be transmitted~too.
\end{definition}

\begin{algorithm}[h!]
\caption{Sampling a Bernoulli random variable with
approximate defining parameter}\label{algo-Bernoulli}
\newcounter{pagealgobernoulli}
\setcounter{pagealgobernoulli}{\thepage}
\addtocounter{pagealgobernoulli}{1}
\begin{algorithmic}[1]
\raggedright
\STATE Set $k \leftarrow 1$
\STATE Set $U[0] \leftarrow 0$
\LOOP
\STATE Generate an i.i.d.\ unbiased bit $U_k$
\STATE Compute $U[k] \leftarrow U[k-1] + U_k/2^k$ \{hence $U[k]=0.U_1\ldots U_k$\}
\STATE Obtain $p[k]$ so that $|p[k]-p|\leq 1/2^k$
\IF{$U[k] \le p[k] - 2/2^k$}
\STATE \textbf{return} $Y = 1$
\ELSIF{$U[k] \geq  p[k] + 1/2^k$}
\STATE \textbf{return} $Y = 0$
\ELSE
\STATE $k \leftarrow k+1$
\ENDIF
\ENDLOOP
\end{algorithmic}
\end{algorithm}

We~postpone to Section~\ref{Bernoulli-distributed} the detail of how these approximations
can be obtained in a distributed
setting. For the moment, assume that we can obtain $p[k]$ so that \mbox{$|p[k]-p|\leq 1/2^k$} for any~$k$.
Then, setting $U[k] = 0.U_1\ldots U_k$,
we have that $U \le p$ if \mbox{$U[k] \le p[k] - 2/2^k$} whereas $U \ge p$ if $U[k] \ge p[k] + 1/ 2^k$.
Thus, one can check if \mbox{$U < p$}
(again disregarding the probability~$0$ event that \mbox{$U=p$})
by generating only as many bits of $U$ and increasingly good approximations of $p$ as needed.
These ideas are formalized in Algorithm~\ref{algo-Bernoulli} (on~page~\thepagealgobernoulli).
It~is elementary to verify that the~$Y$ generated by this algorithm is Bernoulli$(p)$, again
because \mbox{$\mathbf{P}\{U<p\}=p$} if $U$ is a continuous uniform random variable on~\mbox{$[0,1]$}.

The number of iterations before Algorithm~\ref{algo-Bernoulli} returns a value,
which is also its required number of independent unbiased random bits, is a random variable, say~$K$\@.
We~have seen above that $\mathbf{E} \{ K \}$, the expected value of~$K$, would be exactly~2 if we could generate
arbitrarily precise truncations of~$p$. But since we can only obtain arbitrarily precise approximations instead,
which is why we needed Algorithm~\ref{algo-Bernoulli} in the first place, we shall have to pay the price of a small increase in~$\mathbf{E} \{ K \}$.
\begin{displaymath}
\mathbf{P} \{ K > k \}
\le \mathbf{P} \left\{ | U[k]-p[k] | \le \frac{2}{2^k} \right\}
\le\mathbf{P} \left\{ | U-p | \le \frac{4}{2^k} \right\}
\le \frac{8}{2^k }.
\end{displaymath}
Therefore,
\begin{displaymath}
\mathbf{E} \{ K \} = \sum_{k=0}^\infty \mathbf{P} \{ K > k \} \le \sum_{k=0}^\infty \min \left( 1 , \frac{8}{2^k } \right) = 5.
\end{displaymath}

\subsection{Sampling \boldmath{$p_1$} (or \boldmath{$p_2$}) in the random bit model}\label{sampling_p}
As mentioned already, it suffices to concentrate on $p_1$ since one can sample $p_2$
in exactly the same way provided one of the angles $\varphi_j$ is replaced by $\varphi_j+2\pi$:
this introduces the required minus sign in front of $\mathrm{a}_2$ to transform $p_1$ into~$p_2$.
Let us define
\begin{equation}\label{eq:alphabeta}
\textstyle \alpha_j = \cos\!\big(\frac{1}{2}\big(\varphi_j-\frac{\pi}{2}\big)\big)
                             = \sin\!\big(\frac{1}{2}\big(\varphi_j+\frac{\pi}{2}\big)\big)
  \mbox{~~~and~~~}
               \beta_j = \cos\!\big(\frac{1}{2}\big(\varphi_j+\frac{\pi}{2}\big)\big)
                           = -\sin\!\big(\frac{1}{2}\big(\varphi_j-\frac{\pi}{2}\big)\big)  \, .
\end{equation}
Clearly, $\alpha_j^2 + \beta_j^2 = 1$.
Now, consider $n$ Rademacher\,\footnote{\,A Rademacher random variable
is just like a Bernoulli, except that it takes value $+1$ or $-1$, 
rather than $0$ or~$1$.}
random variables $B_j$ that take value $-1$
with probability $\beta_j^2$ and $+1$ with {complementary }probability~$\alpha_j^2$.
The random vector with independent components given by \mbox{$(B_1 , \ldots , B_n)$} is distributed
according to
\begin{displaymath}
q_1(b) \stackrel{\textrm{def}}{=} \prod_{j \in F_b} \beta_j^2 \prod_{j \in G_b} \alpha_j^2 \, ,
\end{displaymath}
where
\mbox{$F_b = \{ j \mid b_j=-1 \}$} and \mbox{$G_b = \{ j \mid b_j=+1 \}$}
for all \mbox{$b = (b_1, \ldots, b_n) \in \{ -1,+1 \}^n$}.
It~is easy to verify that \mbox{$q_1(b) = \mathrm{a}_1^2(b)$} for all~$b$,
where $\mathrm{a}_1$ is given in Equation~(\ref{eq:a1a2}).
Similarly, the random vector with independent components given by \mbox{$(-B_1 , \ldots , -B_n)$} is distributed
according to
\begin{displaymath}
q_2(b) \stackrel{\textrm{def}}{=} \prod_{j \in F_b} \alpha_j^2 \prod_{j \in G_b} \beta_j^2 = \mathrm{a}_2^2(b)\,.
\end{displaymath}

The key observation is that both $q_1$ and $q_2$ can be sampled without any needs for communication
because each party $j$ knows his own parameters $\alpha_j^2$ and~$\beta_j^2$,
which is sufficient to draw independently according to local Rademacher random variable $B_j$ or~$-B_j$.
\mbox{Moreover}, a single unbiased independent random bit~$S$ drawn by a designated party suffices
to sample collectively from distribution \mbox{$q=\frac{q_1+q_2}{2}$}, provided this bit is transmitted
to all parties: everybody samples according to $q_1$ if \mbox{$S=0$} or to $q_2$ if \mbox{$S=1$}.
Now, It follows from Equation~(\ref{eq:p1p2}) that
\mbox{$p_1(b)+p_2(b) = \mathrm{a}_1^2(b)+\mathrm{a}_2^2(b) = q_1(b)+q_2(b)$}
for all \mbox{$b\in\{ -1,+1 \}^n$}, and therefore \mbox{$p_1(b) \le q_1(b)+q_2(b) = 2q(b)$}.

The relevance of all these observations is that we can apply
von Neumann's rejection algorithm~\cite{vN51} to sample $p_1$ since
it is bounded by a small constant~(2) times an easy-to-draw probability distribution~($q$).
For~the moment, we assume once again the availability of a continuous uniform random generator,
which we shall later replace by a source of unbiased independent random bits.
We~also assume for the moment that we can compute the $\alpha_j$'s, $p_1(b)$, $q_1(b)$ and $q_2(b)$
exactly.
This gives rise to Algorithm~\ref{algo-vN}.

\begin{algorithm}[h!]
\caption{Sampling $p_1$ using von Neumann's rejection algorithm}\label{algo-vN}
\begin{algorithmic}[1]
\raggedright
\REPEAT
\STATE   Generate $U$ uniformly on $[0,1]$
\STATE   Generate independent Rademacher random variables $B_1,\ldots,B_n$ \\
               with parameters $\alpha_1^2,\ldots,\alpha_n^2$
\STATE   Generate an unbiased independent random bit $S$
\IF{ $S=1$ }
\STATE set $B \leftarrow (B_1, \ldots, B_n)$
\ELSE
\STATE set $B \leftarrow (-B_1, \ldots, -B_n)$
\ENDIF
\UNTIL{ $U \, (q_1(B)+q_2(B)) \le p_1(B)$}
\end{algorithmic}
\end{algorithm}

By the general principle of von Neumann's rejection algorithm,
probability distribution $p_1$ is successfully sampled after an expected number of 2 iterations round the loop
because \mbox{$p_1(b) \le 2q(b)$} for all $b\in\{-1,+1\}^n$.
Within one iteration, 2 expected independent unbiased random bits suffice to generate each of
the $n$ Rademacher random variables by a process similar to what is explained in the second
paragraph of Section~\ref{Bernoulli}. Hence an expected total of \mbox{$2n+1$} random bits are needed
each time round the loop for an expected grand total of \mbox{$4n+2$} bits
to sample~$p_1$.
But of course, this does not take account of the (apparent) need to generate the
continuous uniform $[0,1]$ random variable~$U$\@.
It~follows that the expected total amount of work required by Algorithm~\ref{algo-vN} is $O(n)$,
provided we count infinite real arithmetic at unit cost.
Furthermore,
the time taken by this algorithm,
divided by~$n$, is stochastically smaller
than a geometric random variable with constant mean, so its tail is
exponentially decreasing.

Now, we modify and adapt this algorithm to eliminate the need for the continuous uniform~$U$
(and hence its generation), which is not allowed in the random \emph{bit} model.
Furthermore, we eliminate the need for infinite real arithmetic and for
the exact values of $q_1(B)$, $q_2(B)$ and $p_1(B)$,
which would be impossible to obtain in our distributed setting since the parameters needed
to compute these values are scattered among all parties,
and replace them with approximations---we~postpone to Section~\ref{comcomp}
the issue of how these approximations can be computed.
(On~the other hand, arbitrarily precise values of the $\alpha_j$'s \emph{are} available
to generate independent Rademacher random variables with these
parameters because each party will be individually responsible to generate his own Rademacher.)

In~each iteration of Algorithm~\ref{algo-vN}, we generated
a pair \mbox{$(U, B)$}.
However, we did not really need~$U$:
we~merely needed to generate a Bernoulli random variable $Y$
for which
\begin{displaymath}
\mathbf{P} \{ Y=1 \} = \mathbf{P} \left\{ U \, (q_1(B)+q_2(B))  \le p_1(B)   \right\}.
\end{displaymath}
For this, we adapt the method developed for Algorithm~\ref{algo-Bernoulli}.
Again, we denote by $U[k]$ the \mbox{$k$-bit} truncation of $U$,
so that \mbox{$U[k] \le U \le U[k]+2^{-k}$}.
Furthermore, we use $L_k$
($L$~for \emph{left}) and $R_k$ ($R$~for \emph{right}) to denote \mbox{$k$-bit} approximations of $q_1(B)+q_2(B)$ and $p_1(B)$, respectively, so that \mbox{$|L_k-\big(q_1(B)+q_2(B)\big)|\leq 2^{-k}$} and $|R_k-p_1(B)|\leq 2^{-k}$.
Then, we use $\varepsilon_k$
to denote the real number in interval~\mbox{$[-1,1]$} so that
\begin{eqnarray*}
| U[k]L_k - U \, (q_1(B)+q_2(B)) |&=& \Big| U[k]L_k - U \, \Big(L_k+\frac{\varepsilon_k}{2^{k}}\Big)\Big| \\
&= &  \Big|(U[k]-U)L_k-\frac{U\varepsilon_k}{2^{k}}\Big|
~\leq~ \frac{L_k}{2^{k}}+\frac{1}{2^{k}}
~\leq~ \frac{3}{2^{k}}\,.
\end{eqnarray*}
Furthermore, because $R_k$ is a \mbox{$k$-bit} approximation of~$p_1(B)$,
\begin{displaymath}
| R_k - p_1(B) | \le \frac{1}{2^k} \,.
\end{displaymath}
Thus, we know that $Y=1$ whenever
\begin{displaymath}
U[k]L_k + 3 / 2^k  < R_k - 1 / 2^k \, ,
\end{displaymath}
whereas $Y=0$ whenever
\begin{displaymath}
U[k]L_k - 3 / 2^k > R_k + 1 / 2^k \, .
\end{displaymath}
Otherwise, we are in the uncertainty zone and we need more bits of $U$, \mbox{$q_1(B)+q_2(B)$}
and $p_1(B)$
before we can decide on the value of~$Y\!$.
This is formalized in Algorithm~\ref{algo-stopping}.

\begin{algorithm}
\caption{Generator for the stopping condition in Algorithm~\ref{algo-vN}}\label{algo-stopping}
\begin{algorithmic}[1]
\raggedright
\STATE Note: $B\in \{ -1,+1\}^n$ is given to the algorithm, generated according to $\frac{q_1+q_2}{2}$
\STATE Set $k \leftarrow 1$
\STATE Set $U[0] \leftarrow 0$
\LOOP
\STATE Generate an i.i.d.\ unbiased bit $U_k$
\STATE Compute $U[k] \leftarrow U[k-1] + U_k/2^k$ \{hence $U[k]=0.U_1\ldots U_k$\}
\STATE\label{step:computeLkRk} Compute $L_k$ and $R_k$ from $B$
\IF {$U[k]\,L_k - R_k < - \frac{4}{2^k}$}
\STATE \textbf{return} $Y = 1$
\ELSIF{$U[k]\,L_k - R_k >  \frac{4}{2^k}$}
\STATE \textbf{return} $Y = 0$
\ELSE
\STATE $k \leftarrow k+1$
\ENDIF
\ENDLOOP
\end{algorithmic}
\end{algorithm}

It~follows from the above discussion that this algorithm can be used to sample random variable~$Y\!$,
which is used as terminating condition in Algorithm~\ref{algo-vN}, in order
to eliminate the need for the generation of a continuous uniform random variable
\mbox{$U \in [0,1]$} and for the precise values of $q_1(B)$, $q_2(B)$ and~$p_1(B)$.
Since $L_k \to q_1(B)+q_2(B)$ and $R_k \to p_1(B)$ as $k \rightarrow \infty$,
Algorithm~\ref{algo-stopping}  halts with probability~1.
Let $K$ be a random variable corresponding to the value of $k$ upon exiting from the
\textbf{loop} in the
algorithm, which is the number of times round the loop and hence the number of bits needed from $U$ and the precision in \mbox{$q_1(B)+q_2(B)$} and $p_1(B)$ required in order to sample correctly Bernoulli random variable~$Y\!$\@.
Next, we calculate an upper bound on $\mathbf{E} \{ K \}$, the expected value of~$K$.

If~the algorithm has not yet halted after having processed $U[k]$, $L_k$ and $R_k$, then we know that
\begin{eqnarray*}
 && |U \, (q_1(B)+q_2(B))-p_1(B)| \\[1ex]
 && ~~~~~~~~=~~\big|\big(U \, (q_1(B)+q_2(B))-U[k]L_k\big)+\big(R_k-p_1(B)\big)+\big(-R_k+U[k]L_k\big)\big|\\[1ex]
 && ~~~~~~~~\leq~~|U \, (q_1(B)+q_2(B))-U[k]L_k|+|R_k-p_1(B)|+|R_k-U[k]L_k|\\[1ex]
 && ~~~~~~~~\leq~~\frac{3}{2^{k}}+\frac{1}{2^{k}}+\frac{4}{2^{k}}
 ~~=~~\frac{8}{2^{k}}\,.
\end{eqnarray*}
Therefore
\begin{align*}
\mathbf{P}\{K>k\mid B\}&\leq \mathbf{P}\{|U \, (q_1(B)+q_2(B))-p_1(B)|\leq 8\slash2^{k}\mid B\}\\[1ex]
&=\mathbf{P}\left\{U \in \left( \frac{p_1(B)}{2q(B)}-\frac12 \frac{8}{2^{k}}\frac{1}{q(B)} ~,~ \frac{p_1(B)}{2q(B)}+\frac12\frac{8}{2^{k}}\frac{1}{q(B)} \right) \right\} \\[1ex]
&\le \frac{8}{2^{k}}\frac{1}{q(B)}\,.
\end{align*}
Thus, using $k_0$ to denote \mbox{$3+\lceil \, \log_2 (1/q(B)) \rceil $},
\begin{eqnarray*}
\mathbf{E} \{ K \mid B \} &=& \sum_{k=0}^\infty \mathbf{P} \{ K > k \mid B \} \\
&\le & \sum_{k=0}^\infty \min \left( 1 , \frac{8}{2^k q(B) } \right) \\
&\le & \sum_{k < k_0} 1  + \,\, \sum_{k \ge k_0} \frac{8}{2^k q(B) } \\[1ex]
&\le & 5 + \log_2 \left(\frac{1}{q(B)} \right)\, .
\end{eqnarray*}
\fussy
The last step uses the fact that \mbox{$x+2^{1-x} \le 2$} for all
\mbox{$0 \le x < 1$}, where \mbox{$x=\lceil \, \log_2 (1/q(B)) \rceil  - \log_2 (1/q(B))$}.

\sloppy
Finally, we uncondition in order to conclude:
\begin{eqnarray}
\mathbf{E} \{ K \}  &\le  & 5 + \sum_{b \in \{-1,+1\}^n} {q(b)} \log_2 \left( \frac{1}{q(b) } \right) \nonumber \\[1ex]
&=& H(q) + 5  \label{eq:H} \\
&\le &n + 5\,,  \label{eq:Nplus5}
\end{eqnarray}
where $H(q)$ denotes the Shannon entropy of distribution $q=\frac{q_1+q_2}{2}$.

\section{Communication complexity of sampling}\label{comcomp}
In this section, we consider the case in which the sampler of the previous section no longer
has full knowledge of the GHZ distribution to be simulated. The sampler, whom we call
\emph{the leader} in a distributed setting, has to communicate through classical channels
in order to obtain partial knowledge of the parameters belonging to the other parties.
Partial knowledge results in approximations of the parameters involved in sampling
the GHZ distribution, but, as we saw in the previous section, we know how to sample
\emph{exactly} in the random bit model using such approximations.
We~consider two models of communication: in~the \emph{sequential model},
the leader has a direct channel with everyone else and all the communication has to take place sequentially
because the leader cannot listen to everyone at the same time;
in~the \emph{parallel model}, parties communicate with one another in a tree-structured way,
with the leader at the root, which makes it possible to save on communication \emph{time},
at the expense of a small increase in the total number of bits that need to be communicated.
Unless specified otherwise, mostly in Section~\ref{parallel}, the sequential model is implicitly assumed.

\subsection{Approximating sums and products of bounded numbers}\label{sct:sums-and-products}
We shall need to approximate sums and products of numbers for which we already
have approximations or truncations.
\begin{theorem}\label{thm:sums-and-products}
Let $k$ and $v$ be integers and
consider any two real numbers $x$ and $y$ in interval $[-2^v,2^v]$.
Let~$\hat{x}$ and~$\hat{y}$ be arbitrary \mbox{$k$-bit} approximations
of $x$ and $y$, respectively,
also restricted to lie in interval $[-2^v,2^v]$.\,
\footnote{\,In case $\hat{x}$ and/or $\hat{y}$ would lie slightly outside $[-2^v,2^v]$,
they can be 
pushed back on the frontier of this interval.}
Then,
\begin{enumerate}
\item $\hat{x}+\hat{y}$ is a \mbox{$(k-1)$-bit} approximation of $x+y$;
\item $\hat{x}/2$ is a \mbox{$(k+1)$-bit} approximation of $x/2$; and
\item $\hat{x}\hat{y}$ is a \mbox{$(k-v-1)$-bit} approximation of $xy$.
\end{enumerate}
\end{theorem}
\begin{proof}
\begin{enumerate}
\item $| (\hat{x}+\hat{y})-(x+y) | = | (\hat{x}-x)+(\hat{y}-y) | \le | (\hat{x}-x) | + | (\hat{y}-y) |
\le 2^{-k}+2^{-k} = 2^{-(k-1)}$;
\item $| \frac{\hat{x}}{2} - \frac{x}{2} | = \frac{| \hat{x} - x |}{2} \le 2^{-k}/2 = 2^{-(k+1)}$; and
\item $|xy - \hat{x}\hat{y}| = \frac12 \, \big|(x+\hat{x})(y-\hat{y}) + (x-\hat{x})(y+\hat{y})\big|
\le \frac12 \big( |(x+\hat{x})(y-\hat{y})| + |(x-\hat{x})(y+\hat{y})| \big) $\\*
$\phantom{|xy - \hat{x}\hat{y}|} \le \frac12 \big(  (| x | + | \hat{x} |) 2^{-k} + 2^{-k} (| y | + | \hat{y} | ) \big)
\le \frac12 \big( (2^v + 2^v)2^{-k} + 2^{-k}(2^v + 2^v) \big) = 2^{-(k-v-1)}$,
\end{enumerate}
\noindent where we used throughout the triangle inequality \mbox{$|a+b| \le |a|+|b|$}.
\end{proof}
\noindent
\begin{corollary}\label{cor:sums-and-squares}
Let $k$, $v$, $x$, $y$, $\hat{x}$ and~$\hat{y}$ be as in Theorem~\ref{thm:sums-and-products}.
\begin{enumerate}
\item $\hat{x}^2+\hat{y}^2$ is a \mbox{$(k-v-2)$-bit} approximation of $x^2+y^2$; and
\item $\frac12 (\hat{x}+\hat{y})^2$ is a \mbox{$(k-v-2)$-bit} approximation of $\frac12 (x+y)^2$.
\end{enumerate}
\end{corollary}
\begin{proof}
This follows from Theorem~\ref{thm:sums-and-products}, using
the fact that the sum of two numbers in interval \mbox{$[-2^v,2^v]$}
lies in interval \mbox{$[-2^{v+1},2^{v+1}]$}.
\end{proof}

\begin{corollary}\label{cor:iterated-products}
Let $k$ and \mbox{$n > 2^k$} be integers and let $\{x_j\}_{j=1}^n$ and $\{\hat{x}_j\}_{j=1}^n$
be real numbers and their \mbox{$k$-bit} approximations, all in interval \mbox{$[-1,1]$}.
Then $\prod_{j=1}^n \hat{x}_j$ is a \mbox{$(k-\lceil \, \log_2 n \rceil)$-bit} approximation
of~$\prod_{j=1}^n x_j$\,.
\end{corollary}
\begin{proof}
Let us place the $\hat{x}_j$'s in the leaves of a binary tree of height $\lceil \, \log_2 n \rceil$.
If~each internal node represents the product of its two children, the corollary follows from
repeated use of Theorem~\ref{thm:sums-and-products}, using \mbox{$v=0$}, since we lose
one bit of precision at each level up the tree until we reach $\prod_{j=1}^n \hat{x}_j$
at the root.
\end{proof}

\subsection{Sampling a Bernoulli distribution whose parameter is distributed}\label{Bernoulli-distributed}
In order to sample the GHZ distribution, we know from Section~\ref{sampling}
that we must first sample the Bernoulli distribution with parameter $\cos^2(\theta/2)$,
where \mbox{$\theta = \sum_{j=1}^{n} \theta_j$}.
Let us say that the leader is party number~1. Since he knows only~$\theta_1$,
he must communicate with the other parties to obtain \mbox{partial} knowledge
about $\theta_j$ for $j \ge 2$. The problem of sampling a Bernoulli distribution
with probability $\cos^2(\theta/2)$ reduces to learning the sum $\theta$ with
sufficient precision in order to use Algorithm~\ref{algo-Bernoulli}.

To compute a \mbox{$k$-bit} approximation of $\cos^2(\theta/2)=\cos^2\!\big(\sum_{j=1}^{n}{\theta_j/2}\big)$, 
define \mbox{$\vartheta=\theta/2$} and \mbox{$\vartheta_j=\theta_j/2$} for each~$j$.
If~the leader obtains an \mbox{$\ell$-bit} approximation \smash{$\hat{\vartheta_j}$} of~each $\vartheta_j$, \mbox{$j \ge 2$}, and if we define
\mbox{\smash{$\hat{\vartheta}=\sum_{j=1}^{n} \hat{\vartheta_j}$}},
we need~to find the value of $\ell$ for which \smash{$\cos^2(\hat{\vartheta})$} is a \mbox{$k$-bit} approximation of \mbox{$\cos^2(\vartheta)$}.
By~virtue of standard results on Taylor series expansion, we have
\begin{displaymath}
|\!\cos^2(\vartheta)-\cos^2(\hat{\vartheta})| \leq \Bigg(\sup_{(\vartheta_1,\ldots,\vartheta_n)}{\|\nabla\big(\cos^2(\vartheta)\big)\|}\Bigg)\|\vartheta-\hat{\vartheta}\| \leq \sqrt{n}\,\frac{\sqrt{n}}{2^{\ell}} = \frac{n}{2^{\ell}} \, ,
\end{displaymath}
where \mbox{$\|\cdot\|$} denotes the Euclidean norm of a vector.
Hence, it suffices to choose \mbox{$\ell = k+\lceil\,\log_2 n\rceil$}
in order to conclude as required that \mbox{\smash{$|\!\cos^2(\vartheta)-\cos^2(\hat{\vartheta})| \leq 2^{-k}$}}.
Taking into account the integer part of each $\vartheta_j$, which must also be communicated,
and remembering that \mbox{$0 \le \vartheta_j \le 2\pi$} since it is an angle\,\footnote{\,Actually, \mbox{$0 \le \vartheta_j \le \pi$} since $\vartheta_j$ is a \emph{half} angle and one fewer bit is needed
to communicate its integer part, but we prefer to consider here the more general case of approximating
the cosine square of a sum of arbitrary angles.},
the required number of communicated bits in the sequential model is therefore
\mbox{$(n-1)(\ell+3)=(n-1)\big(k+3+\lceil\,\log_2 n\rceil \big)$},
which is $O(kn + n \log n)$.
In~our case, the expected value of $k$ is bounded by~$5$
(see the analysis of the Bernoulli sampling Section~\ref{Bernoulli}),
so that this operation requires an expected
communication of $O(n \log n)$ bits.

\subsection{Running von Neumann's rejection algorithm in a distributed setting}\label{approx-product}
Once the leader has produced a bit $Z$ according to a Bernoulli distribution with parameter
$\cos^2(\theta/2)$, he samples either $p_1$ or~$p_2$, depending on whether
he got \mbox{$Z=0$} or~\mbox{$Z=1$}. The problem of sampling $p_2$ \mbox{reduces}
to sampling $p_1$ if the leader replaces his own $\varphi_1$ with $\varphi_1+2\pi$;
thus we concentrate on sampling $p_1$. Of~course, the leader does not know
$\varphi_j$ for $j \ge 2$. In~order to apply von Neumann's rejection method
from Section~\ref{sampling_p}
(Algorithms~\ref{algo-vN} and~\ref{algo-stopping}), the leader needs the ability to learn
with sufficient precision the products \smash{\mbox{$\mathrm{a}_1(B)=\prod_{j=1}^{n} c_j$}}
and \smash{\mbox{$\mathrm{a}_2(B)=\prod_{j=1}^{n} s_j$}}, where
\smash{\mbox{$c_j=\cos\big(\frac{1}{2}\big(\varphi_j-\frac{\pi}{2}B_j\big)\big)$}} and
\smash{\mbox{$s_j=-\sin\big(\frac{1}{2}\big(\varphi_j-\frac{\pi}{2}B_j\big)\big)$}},
given that~the $B_j$'s are non-identical independent Rademacher distributions with parameters
$\alpha_j^2$, \mbox{$1 \le j \le n$}, defined in Equation~(\ref{eq:alphabeta}).
Once these products
are known with $k+2$ bits of precision, the left and right \mbox{$k$-bit} approximations $L_k$ and $R_k$ are easily computed by virtue of Corollary~\ref{cor:sums-and-squares}, using \mbox{$v=0$}.
This is the information needed at line~\ref{step:computeLkRk} of Algorithm~\ref{algo-stopping}.

It~follows from Corollary~\ref{cor:iterated-products} that if we set
\mbox{$\ell=k+2+\lceil \, \log_2 n \rceil$} and if the leader obtains \mbox{$\ell$-bit}
approximations $\hat{c}_j$ and $\hat{s}_j$ of each party's $c_j$ and~$s_j$,
he can compute the required  \mbox{$(k+2)$-bit} approximations of
$\mathrm{a}_1(B)$ and~$\mathrm{a}_2(B)$.
\emph{Notice that each party knows exactly his own $c_j$
and $s_j$,
and hence $\hat{c}_j$ and $\hat{s}_j$ can be transmitted directly
to the leader, rather than approximations of the $\varphi_j$'s}.
These \mbox{$\ell$-bit} approximations can in fact be \mbox{$\ell$-bit} truncations,
requiring the transmission of \mbox{$\ell+1$} bits, taking account of the sign,
for a grand total of $2(n-1)(k+3+\lceil \, \log_2 n \rceil)$ bits that must be transmitted
to the leader, which is \mbox{$O(kn+n \log n)$}.
For our specific application of simulating the GHZ distribution, we proved
at the end of Section~\ref{sampling_p} (Equation~\ref{eq:Nplus5})
that the expected value of $k$ is bounded by~\mbox{$n+5$}.
It~follows that an expected communication cost of $O(n^2)$ bits suffices to
sample the GHZ distribution, as we shall prove formally in the next section.

\section{Protocol for sampling the GHZ distribution}\label{final-protocol}
\begin{algorithm}
\caption{Complete protocol for sampling the GHZ distribution in the sequential model}\label{algo-final}
\newcounter{pagealgo}
\setcounter{pagealgo}{\thepage}
\addtocounter{pagealgo}{1}
\begin{algorithmic}[1]
\raggedright
\STATE The leader, who is party number~1, communicates with the other parties in order to obtain increasingly precise approximations of \mbox{$\theta=\sum_{j=1}^{n}\theta_j$} until he can sample random bit $Z$ according to \emph{exact} Bernoulli random distribution with parameter $\cos^2(\theta/2)$\label{step:Bern}
\IF{$Z=1$}
\STATE The leader adds $2\pi$ to his own $\varphi$-parameter, i.e.~$\varphi_1 \leftarrow \varphi_1 + 2\pi$ \\
\COMMENT{to sample $p_2$ rather than~$p_1$}
\ENDIF\\[1ex]
\COMMENT{Now entering von Neumann's sampling algorithm for $p_1$, adapted to our distributed setting}
\REPEAT\label{begin-repeat}
\STATE The leader generates a fair random bit $S$ and broadcasts it to the other parties \\
\COMMENT{The bit $S$ determines whether to sample $q_1$ or $q_2$}\label{step:scratch}
\STATE Locally, each party $j$ generates a random $B_j \in \{-1,+1\}$ according to an independent Rademacher distribution
so that \mbox{$B_j=+1$} with probability $\cos^2\!\big(\frac{1}{2}\big(\varphi_j-\frac{\pi}{2}\big)\big)$ \\
\COMMENT{Random variable $B=(B_1,\ldots,B_n)$ is now sampled according to~$q_1$}\\
\IF {$S=1$}
\STATE Each party does $B_j \leftarrow -B_j$\\
\COMMENT{In this case, random variable $B=(B_1,\ldots,B_n)$ is now sampled according to~$q_2$}\\
\ENDIF\\
\COMMENT{Random variable $B=(B_1,\ldots,B_n)$ is sampled according to~$q=\frac{q_1+q_2}{2}$}\\[1ex]
\COMMENT{The leader starts talking
with the other parties to decide whether or not to accept~$B$}
\STATE{Each party computes $c_j=\cos\!\big(\frac{1}{2}\big(\varphi_j-\frac{\pi}{2}B_j\big)\big)$ and
$s_j=-\sin\!\big(\frac{1}{2}\big(\varphi_j-\frac{\pi}{2}B_j\big)\big)$}\\
\STATE The leader sets $k \leftarrow 1$\label{line:initk}
\STATE The leader sets $U[0] \leftarrow 0$ \\
\LOOP\label{begin-repeat-forever}
\STATE The leader generates an i.i.d.\ unbiased bit $U_k$
\STATE The leader computes $U[k] \leftarrow U[k-1] + U_k/2^k$ \{hence $U[k]=0.U_1\ldots U_k$\}
\STATE The leader requests $(k+2+\lceil \, \log_2 n \rceil)$-bit truncations of $c_j$ and $s_j$ from each party $j \ge 2$\label{line:costly}
\STATE The leader computes $(k+2)$-bit approximations of $\mathrm{a}_1(B)$ and $\mathrm{a}_2(B)$\label{line:costly-bis}
\STATE The leader computes \mbox{$k$-bit} approximations $L_{k}$ of $\mathrm{a}_1^2(B) + \mathrm{a}_2^2(B)$ and $R_{k}$ of
$p_1(B)$
\IF {$U[k] \, L_k - R_k < - \frac{4}{2^k}$}
\STATE Set $Y \leftarrow 1$ and \textbf{break from the loop}. \COMMENT{Vector $B$ is accepted}
\ELSIF{$U[k] \, L_k - R_k >  \frac{4}{2^k}$}
\STATE Set $Y \leftarrow 0$ and \textbf{break from the loop}. \COMMENT{Vector $B$ is rejected}
\ELSE
\STATE Set $k \leftarrow k+1$ and \textbf{continue the loop} \label{line:increase} \\
\COMMENT{The leader does not yet have enough information to decide whether to accept or \\
~~reject~$B$; therefore, he needs more information from all the other parties in order \\
~~to compute one more bit of precision on $\mathrm{a}_1(B)$ and $\mathrm{a}_2(B)$}
\ENDIF
\ENDLOOP\label{end-repeat-forever}
\UNTIL{$Y=1$ \COMMENT{accepting}}\label{end-repeat}
\STATE The leader informs the other parties that the simulation is complete and, therefore, that the time has
come for each party~$j$ (including the leader himself) to output his current value of~$B_j$\label{step:thatsit}
\end{algorithmic}
\end{algorithm}

We are finally ready to glue all the pieces together into Algorithm~\ref{algo-final}
(on~page~\thepagealgo),
which samples exactly the GHZ distribution
under arbitrary von Neumann measurements, thus solving our conundrum.
Its correctness is proved below, and it is shown that
the expected amount of randomness
used in this process is upper-bounded by $O(n)$ bits and an expected $O(n^2)$ bits
of communication suffice to complete the task.
Fewer bits suffice when the measurements are in the computational basis or nearly~so.

\subsection{Correctness of the protocol}
The part occurring before line~\ref{begin-repeat} samples a Bernoulli with parameter
$\cos^2\!\big(\!\sum_{j=1}^{n}{\theta_j/2}\big)$, which allows the leader to decide whether
to sample $B$ \mbox{according} to~$p_1$
(by~leaving his $\varphi_1$ unchanged) or according to~$p_2$ (by~adding $2\pi$ to his~$\varphi_1$).
Notice that the leader does not have to inform the other parties of this decision
since they do not need to know if the sampling will be done according to $p_1$ or~$p_2$.
In~Section~\ref{Bernoulli-distributed}, we showed how to sample exactly a Bernoulli
with parameter $\cos^2\!\big(\!\sum_{j=1}^{n}{\theta_j/2}\big)$ even when the
$\theta_j$'s are not known to the leader for~\mbox{$j \ge 2$}.

The part within the outer \textbf{repeat} loop (lines~\ref{begin-repeat} to~\ref{end-repeat})
is essentially von Neumann's rejection algorithm, which has been
adapted and modified to work in a distributed scenario.
The leader must first decide which of $q_1$ or $q_2$ to sample.
For~this purpose, he generates an \mbox{unbiased} random bit $S$ and broadcasts it to the other parties.
Sampling either $q_1$ or $q_2$ can now be done locally and independently by each party $j$,
yielding a tentative \mbox{$B_j \in \{-1,+1\}$}.
The parties will output these $B_j$'s only at the end, provided this round is not rejected.
Now, each party uses his $B_j$ to compute locally
\smash{\mbox{$c_j=\cos\!\big(\frac{1}{2}\big(\varphi_j-\frac{\pi}{2}B_j\big)\big)$}} and
\smash{\mbox{$s_j=-\sin\!\big(\frac{1}{2}\big(\varphi_j-\frac{\pi}{2}B_j\big)\big)$}},
which will be sent bit by bit to the leader upon request,
thus allowing him to compute increasingly precise approximations $L_k$ and $R_k$ of
\mbox{$q_1(B) + q_2(B)$} and $p_1(B)$, respectively.
These values are used to determine whether a decision can be made to accept or reject this particular~$B$,
or whether more information is needed to make this decision.
As~shown at the end of Section~\ref{sampling_p} (Equation~\ref{eq:Nplus5}),
the expected number of bits needed in $L_k$ and $R_k$
before we can break out of the inner \textbf{loop} (lines~\ref{begin-repeat-forever}
to~\ref{end-repeat-forever}) is \mbox{$k \le n+5$}.
At~that point, flag $Y$ tells the leader whether or not this was a successful
run of von Neumann's rejection algorithm.
If~$Y=0$, the entire process has to be restarted from scratch,
except for the initial Bernoulli sampling, at line~\ref{step:scratch}.
On~the other hand, once the leader gets $Y=1$, he can finally tell the other parties
that they can output their~$B_j$'s because, according to von Neumann's rejection algorithm,
this signals that the vector $(B_1,\ldots,B_n)$ is distributed according to $p_1$
(or~$p_2$, depending on the initial Bernoulli).
Also~according to von Neumann's rejection algorithm, we have an expectation of
\mbox{$C=2$} rounds of the outer \textbf{repeat} loop before we can thus conclude successfully.

\subsection{Expected number of random coins and communication cost}\label{sct:analysis}
The expected amount of randomness used in this process is upper-bounded by \mbox{$6n+17$} bits.
This is calculated as follows: the expected number of bits for sampling Bernoulli~$Z$ is bounded by~$5$.
This is followed by an expectation of \mbox{$C=2$} rounds of von Neumann's rejection algorithm
(the~outer \textbf{repeat}~loop).
In~each of these rounds, we need  $1$ bit for $S$ and expect $2$ bits for each of the $B_j$'s (hence
\mbox{$1+2n$} in total),
before entering the inner \textbf{loop}.
The expected number of times round this loop is bounded by \mbox{$n+5$},
and one more random bit $U_k$ is needed each time.
Putting it all together, the expected number of random bits is bounded by \mbox{$5+2(1+2n+(n+5))=6n+17$}.

The expected amount of communication is dominated by the leader's need to obtain increasingly
accurate approximations of $c_j$ and $s_j$ from all other parties at line~\ref{line:costly}
in \mbox{order} to compute increasingly accurate approximations $L_k$ and $R_k$, which he needs
in \mbox{order} to decide whether or not to break from the inner \textbf{loop} and, in such case,
whether or not to accept~$B$ as final output.
On~the $k^\mathrm{th}$ time round the loop,
the leader needs \mbox{$k+2+\lceil \, \log_2 n \rceil$} bits of precision plus one bit of sign
about each $c_j$ and~$s_j$, \mbox{$j\ge 2$}, in~\mbox{addition} to having full knowledge about his own
$c_1$ and~$s_1$.
\mbox{According} to Section~\ref{approx-product}, this suffices for the leader to compute
\mbox{$(k+2)$-bit} approximations of $\mathrm{a}_1(B)$ and $\mathrm{a}_2(B)$, which in turn
suffice by virtue of Corollary~\ref{cor:sums-and-squares} to obtain
\mbox{$k$-bit} approximations $L_{k}$ of \mbox{$\mathrm{a}_1^2(B) + \mathrm{a}_2^2(B)$} and $R_{k}$ of
\mbox{$p_1(B)=\frac{1}{2}\smash{\big(}\mathrm{a}_1(b)+\mathrm{a}_2(b)\smash{\big)}^2$}.
The need for the leader to obtain from the other parties these increasingly precise
approximations of $c_j$ and~$s_j$
would be very \mbox{expensive} if all their bits had to be resent each time round the loop,
with increasing values of~$k$.
Fortunately, this process works well because the parties actually send \emph{truncations}
of these \mbox{values} to the leader at line~\ref{line:costly}: each truncation simply
adds one bit of precision to the \mbox{previous} one. Hence, it suffices for the leader to request
\mbox{$2(4+\lceil \, \log_2 n \rceil)$} bits from each other party at the onset,
when \mbox{$k=1$}, and only two additional
bits per party are needed afterwards for each subsequent trip round the loop
(one~for $c_j$ and one for~$s_j$).
All counted, a total of \mbox{$2(n-1)(k+3+\lceil \, \log_2 n \rceil)$} bits will have been requested
from all other parties by the time we have gone through the inner \textbf{loop} $k$ times.
Since the expected value of $k$ upon exiting this loop is bounded by \mbox{$n+5$},
the expected number of bits that have to be communicated to the leader
to complete von Neumann's rejection algorithm (lines~\ref{begin-repeat} to~\ref{end-repeat})
is bounded by \mbox{$2(n-1)((n+5)+3+\lceil \, \log_2 n \rceil)$}.
This is $O(n^2)$ expected bits of communication.
The additional amount of communication required to sample Bernoulli~$Z$ at line~\ref{step:Bern}
(which is~\mbox{$(n-1)(8+\lceil\,\log_2 n\rceil)$} bits according to Section~\ref{Bernoulli-distributed}) and for
the leader to broadcast to all parties the value of~$S$, as well as synchronization bits by which he needs
to inform the other parties of \mbox{success} or failure each time round the loop is negligible.
All counted, Algorithm~\ref{algo-final} needs $O(n)$ bits of randomness and $O(n^2)$ bits
of communication in order to sample exactly the GHZ distribution
under arbitrary von Neumann measurements.

The analysis above applies regardless of the set of von Neumann measurements that have to be simulated.
In~some cases, however, it is very pessimistic because the expected number of times round
the inner \textbf{loop} is bounded by
$\mathbf{E} \{ K \}  \le  H(q) + 5$
according to Equation~(\ref{eq:H}),
where $H(q)$ is the entropy of distribution $q=\frac{q_1+q_2}{2}$.
Until now, we had simply used the fact that \mbox{$H(q) \le n$} to conclude that
$\mathbf{E} \{ K \}  \le n+5$, which is Equation~(\ref{eq:Nplus5}).
However, $H(q)$
can be much smaller than $n$ for some~$q$.
In~general,
\[H(q) \le 1+(H(q_1)+H(q_2))/2\]
and
\[H(q_1) = H(q_2) = \sum_{j=1}^n H_2(\alpha_j^2) \, ,\]
where $H_2$ is the binary entropy function and the $\alpha_j$'s are given in Equation~(\ref{eq:alphabeta}).
In~the extreme case of measurements in the computational basis,
which corresponds to \mbox{$\varphi_j=\pm\pi/2$} and hence \mbox{$\alpha_j  \in \{0,1\}$},
we have \mbox{$H_2(\alpha_j^2) = 0$} for all~$j$. It~follows that \mbox{$H(q)=1$},
hence the expected number of times round the inner \textbf{loop} is bounded by~6,
and therefore Algorithm~\ref{algo-final} needs only an expectation of $O(n \log n)$ bits
of communication in order to sample exactly the GHZ distribution
under computational-basis von Neumann measurements.
Of~course, $O(n)$ bits of communication would suffice, even in the worst case,
if we knew ahead of time that all measurements are in the computational basis,
but our protocol works seamlessly with $O(n \log n)$ expected bits of communication even
if the measurements are not \emph{exactly} in the computational basis,
provided $H_2(\alpha_j^2)$ is small enough, and even if
up to $O(\log n)$ of the measurements are arbitrary.
The effect of such measurements on the required expected amount of randomness is less dramatic
since replacing ``\mbox{$n+5$}'' by ``6'' in the analysis above merely reduces the expected number of
random bits from \mbox{$6n+17$} bits to \mbox{$4n+19$}.
In~the case of measurements in the computational basis, however, the parties can sample their
local Rademachers without any need for randomness since they become deterministic.
Hence, provided we modify the protocol accordingly to take account of this special case,
the total expected amount of required randomness is upper-bounded by~$19$ bits.

\section{Variations on the theme}\label{variations}
We can modify Algorithm~\ref{algo-final} in a variety of ways to improve different parameters at the expense
of others. Here, we discuss four of these variations:
the parallel model, bounding the number of rounds, the simulation of equatorial measurements,
and the case in which only the leader has access to a source of randomness.

\subsection{The parallel model}\label{parallel}
Until now, we have concentrated on the
\emph{sequential model} of communication, in which the leader has a direct channel with everyone else
but the other participants do not communicate among themselves.
This forces communication to take place sequentially
because the leader cannot listen to everyone at the same time.
However, as mentioned at the beginning of Section~\ref{comcomp},
it is legitimate to consider a \emph{parallel model},
in which arbitrarily many pairs of parties can communicate simultaneously.
Accordingly, any number of bits can be sent and received in the same time step,
provided no party has to send or receive more than one bit at any given time.
This can reduce considerably the \emph{time} required to complete our task,
without entailing a significant increase in the total \emph{number of bits}
that circulate on the network.

\begin{figure}[h]
\begin{center}
\includegraphics[height = 58.5mm]{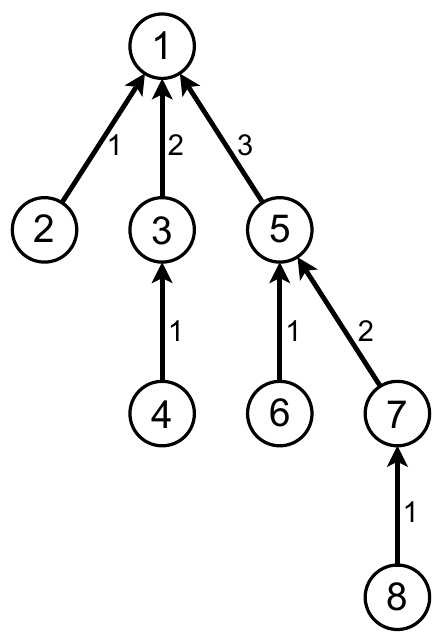}
\end{center}
\caption{Binomial tree structure defining the parallel model.}
\label{binomial}
\end{figure}

In more detail, we have seen in Section~\ref{approx-product} that it is possible for the leader to obtain
\mbox{($k+2$)-bit} approximations of \smash{\mbox{$\mathrm{a}_1(B)=\prod_{j=1}^{n} c_j$}}
and \smash{\mbox{$\mathrm{a}_2(B)=\prod_{j=1}^{n} s_j$}}
at an expected communication cost of \mbox{$O(kn + n \log n)$} bits.
To~bring this down to \mbox{$O(k\log n + \log^2\!n)$} \emph{time},
we let the parties communicate with one another
\mbox{according} to the binomial tree structure shown in Fig.~\ref{binomial},
in which numbers in the nodes correspond to parties (the~leader is number~1 at the root)
and numbers next to the arrows correspond to the order in which data is transmitted.
For~simplicity, we may \mbox{assume} that $n$ is a power of~$2$.
To~understand the algo\-rithm, think of a tree with nodes containing $s_j$ and~$c_j$, for~\mbox{$1 \le j \le n$}. Node number $j$ can be thought of as belonging to party number $j$, who knows $s_j$ and $c_j$ exactly.
We~pair parties by groups of~$2$. For a given pair, say
\mbox{$(j,j+1)$}, with $j$ odd, party \mbox{$j+1$} sends $\hat{s}_{j+1}$ and $\hat{c}_{j+1}$ to party~$j$, who computes $\hat{s}_j \hat{s}_{j+1}$ and $\hat{c}_j \hat{c}_{j+1}$. Then, party~$j$
is matched with party $j+2$,
where \mbox{$j-1$} is divisible by~4. This process gives rise to the new pair \mbox{$(j,j+2)$},
from which emerges the products $\hat{s}_j \hat{s}_{j+1}\hat{s}_{j+2}\hat{s}_{j+3}$ and $\hat{c}_j \hat{c}_{j+1}\hat{c}_{j+2}\hat{c}_{j+3}$, and so on up to the leader, who is party~1 at
the root of the tree.

\begin{algorithm}[h!]
\caption{Computing products in parallel, communicating
along a binomial tree configuration}\label{algo-parallel}
\begin{algorithmic}[1]
\raggedright
\STATE If $n$ is not a power of~$2$,
add virtual parties with \mbox{$c_j=s_j=1$} for
\mbox{$n < j \le 2^{\lceil\,\log_2 n \rceil}$}\\
\COMMENT{These parties, being dummy, have no \mbox{effect} on the overall communication complexity}
\STATE $\ell\leftarrow k+3+\lceil\,\log_2 n\rceil$
\FOR{$j \leftarrow 1$ \textbf{to} $n$ \textbf{in parallel}}
\STATE Party $j$ does $\tilde{c}_j \leftarrow \hat{c}_j$ and $\tilde{s}_j \leftarrow \hat{s}_j$,
which are \mbox{$\ell$-bit} truncations of $c_j$ and $s_j$, respectively.
\ENDFOR
\STATE $m \leftarrow 1$
\REPEAT
\FOR{$j \leftarrow 1$ \textbf{to} $n$
\textbf{by step of} $2m$ \textbf{in parallel}}
\STATE Party $j+m$ sends $\tilde{c}_{j+m}$ and $\tilde{s}_{j+m}$ to party $j$
\STATE Party $j$ computes $\tilde{c}_j \leftarrow \tilde{c}_j \, \tilde{c}_{j+m}$
   and $\tilde{s}_j \leftarrow \tilde{s}_j \, \tilde{s}_{j+m}$, both truncated to $\ell$ bits\label{step:retruncate}
\ENDFOR
\STATE $m \leftarrow 2m$
\UNTIL{$m \ge n$}
\STATE Party 1 (the leader) outputs $\tilde{c}_1$ and $\tilde{s}_1$
\end{algorithmic}
\end{algorithm}

This approach is formalized in Algorithm~\ref{algo-parallel}, in which
new variables $\tilde{c}_j$ and $\tilde{s}_j$ are introduced to hold approximations of products of increasingly
many $c$'s and $s$'s as the process unfolds.
The~issue of the required precision at each level of the process has to be reconsidered
because the leader will no longer receive the entire list of $\hat{c}_j$'s and $\hat{s}_j$'s since subproducts
are calculated \emph{en~route} by intermediate parties, which must be truncated for transmission.
For simplicity, we proceed as if all the $c_j$'s and $s_j$'s were nonnegative and we percolate the
signs up the tree separately. We~know from Theorem~\ref{thm:sums-and-products},
using \mbox{$v=0$}, that
if $\hat{x}$ and $\hat{y}$ are \mbox{$\ell$-bit} approximations (in~particular \mbox{$\ell$-bit} truncations)
of $x$ and $y$, respectively, for arbitrary real numbers $x$ and $y$ in $[0,1]$ and integer~$\ell$,
then $\hat{x}\hat{y}$ is an \mbox{($\ell-1$)-bit} approximation of~$xy$.
However, $\hat{x}\hat{y}$ is \emph{not} in general an \mbox{($\ell-1$)-bit} \emph{truncation} of~$xy$
because it could have up to $2\ell$ bits of precision and we do not want to transmit so many bits
up the binomial tree. There is an apparent problem if we transmit the \mbox{$\ell$-bit} truncation of
$\hat{x}\hat{y}$ instead, as we do indeed in Algorithm~\ref{algo-parallel},
because \emph{it} is an \mbox{($\ell-2$)-bit} approximation of~$xy$,
but not necessarily an \mbox{($\ell-1$)-bit} approximation.
Nevertheless, it is shown in Appendix~\ref{App-error} that
the recursive application of pairwise multiplications followed by truncation to $\ell$ bits
results in the loss of only one bit of precision per subsequent level.
Thus, on the $i^\mathrm{th}$ time round the \mbox{\textbf{repeat}} loop,
\mbox{$1 \le i \le \lceil \, \log_2 n \rceil$},
the numbers calculated at line~\ref{step:retruncate}, even after truncation to $\ell$ bits,
are \mbox{($\ell-i-1$)-bit} approximations of the exact product that they represent
(again, up to sign, which is handled separately).
It~follows that the final numbers computed by the leader when \mbox{$i=\lceil \, \log_2 n \rceil$}
are \mbox{($k+2$)-bit} approximations of the products of all the $c_j$'s and the $s_j$'s, as required, provided
we start with \mbox{$\ell = k+3+\lceil\,\log_2 n\rceil$}.

To analyse the communication complexity of this strategy, we consider that bits sent and received in parallel
between disjoint pairs of parties count as a single time step in the global communication process.
The \textbf{repeat} loop is carried out \mbox{$\lceil\,\log_2 n\rceil$} times.
Each time round this loop, parties transmit in parallel two \mbox{$\ell$-bit} approximations, which require
\mbox{$\ell+1$} bits of communication per active party since signs must also be transmitted.
It~follows immediately that the parallel complexity of Algorithm~\ref{algo-parallel} is
\mbox{$2(\ell+1)\lceil\,\log_2 n\rceil = 2(k + 4 + \lceil\,\log_2 n\rceil)\lceil\,\log_2 n\rceil$}, which is
\mbox{$O(k \log n + \log^2\!n$)}. Therefore, this takes $O(k\log n)$ time provided \mbox{$k>\log_2 n$}.

Now, let us use Algorithm~\ref{algo-parallel} to replace lines~\ref{line:costly} and~\ref{line:costly-bis} in Algorithm~\ref{algo-final}.
This allows the leader to obtain his required \mbox{$(k+2)$-bit} approximations of $\mathrm{a}_1(B)$ and $\mathrm{a}_2(B)$
with no need for him to learn all the \mbox{$(k+2+\lceil \, \log_2 n \rceil)$-bit} truncations of $c_j$ and $s_j$ from each party~$j \ge 2$.
We~have just seen that \mbox{$O(k \log n + \log^2\!n$)} parallel time suffices for this task.
Unfortunately, this improvement is incompatible with the idea of transmitting only one more bit of information
for each $c_j$ and $s_j$ when $k$ is increased by~$1$, which was crucial in the efficiency of the
sequential version of Algorithm~\ref{algo-final} studied in Section~\ref{final-protocol}.
The problem stems from the fact that the \mbox{$\ell$-bit} truncation of the product of the \mbox{$\ell$-bit} truncations
of $x$ and $y$ can be entirely different from the \mbox{$(\ell+1)$-bit} truncation of
the product of the \mbox{$(\ell+1)$-bit} truncations of the same numbers.
This is illustrated with \mbox{$x=0.1111\ldots$} and \mbox{$y=0.1001\ldots$} (in~binary, of course).
If~we take \mbox{$\ell=3$}, the truncations of $x$ and $y$ are $0.111$ and $0.100$,
respectively, whose product is $0.011100$.
In~contrast, with \mbox{$\ell=4$}, the truncations of $x$ and $y$ are $0.1111$ and $0.1001$,
respectively, whose product is $0.10000111$.
We~see that the \mbox{$3$-bit} truncation of the product of the \mbox{$3$-bit} truncations is $0.011$,
whereas the \mbox{$4$-bit} truncation of the product of the \mbox{$4$-bit} truncations is $0.1000$, which are
different on each and every bit of the fractional part!
This demonstrates the fact that the bits going up the binomial tree in Algorithm~\ref{algo-parallel}
can change drastically from one run to the next even if a single bit of precision is added
to all nodes at the bottom level, and therefore that we have to start afresh for each new value of~$k$.
As~a consequence, the use of Algorithm~\ref{algo-parallel} to replace lines~\ref{line:costly} and~\ref{line:costly-bis}
in Algorithm~\ref{algo-final} results in an ``improvement'' in which we expect to have to transmit
$\Omega(n^3)$ bits, taking $\Omega(n^2 \log n)$ parallel time to do~so!

Fortunately, there is an easy cure to this problem, which we only sketch here.
In~addition to using Algorithm~\ref{algo-parallel} to replace lines~\ref{line:costly} and~\ref{line:costly-bis} in Algorithm~\ref{algo-final},
we also change line~\ref{line:increase} from \mbox{``$k \leftarrow k+1$''} to \mbox{``$k \leftarrow 2k$''}.
Even though parties have to transmit  up the binomial tree
the entire \mbox{$(k+3+\lceil \, \log_2 n \rceil)$-bit} truncations of each
$c_j$ and $s_j$ for each new value of~$k$, the work done each time round the loop is roughly equivalent to the
sum of all the work done until then. Since we expect to succeed when $k$ is roughly equal to~$n$,
the expected total parallel time is about twice \mbox{$O(k \log n + \log^2\!n$)} with \mbox{$k \approx n$},
which is simply \mbox{$O(n \log n)$}. The expected total number of bits communicated with this approach
is slightly greater than with Algorithm~\ref{algo-final}, but remains~$O(n^2)$.

\subsection{Reducing the number of rounds}
Algorithm~\ref{algo-final} is efficient in terms of the number of bits of randomness as well as the number
of bits of communication, but it requires an expected $O(n)$ rounds, in which the leader and all other parties
take turn at sending messages. This could be prohibitive if they are far apart and their purpose is to try to convince examiners
that they are actually using true entanglement and quantum processes to produce their joint outputs,
because it would prevent them from responding quickly enough to be credible.
The solution should be rather obvious at this point, and we leave the details to the reader.
If~we change line~\ref{line:increase} from \mbox{``$k \leftarrow k+1$''} to \mbox{``$k \leftarrow 2k$''},
the expected number of rounds is decreased from $O(n)$ to~$O(\log n)$.
If~in addition we start with \mbox{``$k \leftarrow n$''} instead of \mbox{``$k \leftarrow 1$''}
at line~\ref{line:initk}, the expected number of rounds becomes a constant.
(Alternatively, we could start with \mbox{``$k \leftarrow n$''} at line~\ref{line:initk} and
step with \mbox{``$k \leftarrow k+n$''} at line~\ref{line:increase}.)

\subsection{Equatorial measurements}
Recall that equatorial measurements are those for which \mbox{$\varphi_j=0$} for each party~$j$.
In~this case, the leader can sample according to~$p_1$ or $p_2$, without any help or communication
from the other parties, since he has complete knowledge of their vanished elevation angles.
Therefore, he can run lines~\ref{begin-repeat} to~\ref{end-repeat} of Algorithm~\ref{algo-final} all by himself!
However, he still needs to communicate in line~\ref{step:Bern} of Algorithm~\ref{algo-final} in order to
know from which of $p_1$ or $p_2$ to sample.
The~only remaining need for communication occurs in line~\ref{step:thatsit},
which has to be modified from ``The leader informs all the other parties that the simulation is complete''
to ``The leader informs all the other parties of which value of \mbox{$B_j \in \{-1,+1\}$}
he has chosen for them''.

Only line~\ref{step:Bern} requires significant communication since the new line~\ref{step:thatsit}
needs only the transmission of \mbox{$n-1$} bits.
We~have already seen at the end of Section~\ref{Bernoulli-distributed} that
line~\ref{step:Bern}, which is a distributed version of Algorithm~\ref{algo-Bernoulli}, requires an expected
communication of $O(n \log n)$ bits in the sequential model.
This~is therefore the complexity of our simulation, which is
an improvement over the previously best technique known to simulate the GHZ
distribution under arbitrary equatorial von Neumann measurements~\cite{BK12},
which required an expectation of $O(n^2)$ bits of communication.

A~more elegant protocol can be obtained if we use Equation~(\ref{eq:eqp}) at the end of
Appendix~\ref{App-decomp},
which gives us a simplified formula for $p(b)$ in the case of equatorial measurements.
Each party $j$ other than the leader can simply choose an independent unbiased Rademacher
\mbox{$b_j \in \{-1,+1\}$} as final output, without any consideration of his own input~$\theta_j$
nor communication with anyone else, and inform the leader of this choice.
It~simply remains for the leader to choose his own $b_1$ in order to make
\smash{\mbox{$\prod_{j=1}^n b_j$}} equal to $+1$ with probability \smash{$\cos^2(\theta/2)$}
or $-1$ with complementary probability~\smash{$\sin^2(\theta/2)$}.
For this, we still need line~\ref{step:Bern} from Algorithm~\ref{algo-final},
which requires an expected communication of $O(n \log n)$ bits.

To adapt this latter protocol to the parallel model, note that the leader does not need to know
all the $b_j$'s chosen by the other parties since he only needs their product, which is either
$+1$ or~$-1$. It~is elementary to adapt Algorithm~\ref{algo-parallel} in order to percolate this
information to the leader up the binomial tree, at a communication cost of $O(n)$ bits but only $O(\log n)$
parallel time. One~can also adapt Algorithm~\ref{algo-parallel} to work with sums
instead of products, which is the relevant operation to parallelize the distributed version
of Algorithm~\ref{algo-Bernoulli}. Sums and products are similar since if
$\hat{x}$ and $\hat{y}$ are \mbox{$t$-bit} approximations of $x$ and $y$, respectively, for an arbitrary
integer~$t$, then \mbox{$\hat{x}+\hat{y}$} and \mbox{$\hat{x}\hat{y}$} are \mbox{$(t-1)$-bit}
approximations of~\mbox{$x+y$} and~$xy$, respectively,
according to Theorem~\ref{thm:sums-and-products}, using~\mbox{$v=0$}.
However, parallelizing sums is easier than products because
the exact sum \mbox{$\hat{x}+\hat{y}$} can be transmitted
with no more bits of precision than each of $\hat{x}$ and~$\hat{y}$, even though one additional bit
is required to transmit the integer part of the sum,\,
\footnote{\,One may be tempted to prevent the accumulation of large angles
by reducing each sum modulo $2\pi$ before transmission up the binomial tree.
However, this would void the advantage we had reaped from the fact that the
fractional part of the sum of {$t$-bit} truncations (as~opposed to their product)
contains only $t$ bits of precision.}
whereas the product $\hat{x}\hat{y}$ could entail twice as many bits of precision than
each of $\hat{x}$ and~$\hat{y}$
(this~is why we needed Appendix~\ref{App-error}).
In~round $k$ of the \textbf{loop} in Algorithm~\ref{algo-Bernoulli},
the leader needs to obtain a \mbox{$k$-bit} approximation of \smash{$\cos^2(\sum_{j=1}^{n}\theta_j/2)$},
which in turn requires the addition of \mbox{$\ell=k+3+\lceil\,\log_2 n\rceil$} bits
from each of~the \mbox{$n-1$} half-angles \mbox{$\theta_j/2$} owned by the various parties
\mbox{$j \ge 2$}. The binomial tree construction makes it possible to percolate this sum
up to the leader through $\lceil\,\log_2 n\rceil$ levels in which it is sufficient to transmit
\mbox{$\ell+i$} bits up the tree for each partial sum (or~initial half angle) at distance $i$ from the leaves.
The~expected cost before the leader obtains the required \mbox{$k$-bit} approximation of
\smash{$\cos^2(\sum_{j=1}^{n}\theta_j/2)$} is therefore $O((k+ \log n) \, n)$ bits of communications
but only $O((k+ \log n) \log n)$ parallel time.
Using once again the fact that the expected number $k$ of rounds in Algorithm~\ref{algo-Bernoulli}
is bounded by~$5$,
the required Bernoulli variable with parameter \smash{$\cos^2(\sum_{j=1}^{n}\theta_j/2)$}
can be sampled exactly after an expected communication cost of $O(n \log n)$ bits,
as in the sequential model,  but only $O(\log^2\!n)$ parallel time.
This dominates the cost of the parallel implementation of our algorithm in the case
of equatorial measurements.

Note that all this information is sent up the binomial tree towards the leader.
The only information that the leader has to send back down to the other parties,
each time round the \textbf{loop},
serves to notify them of whether or not a more precise approximation of their azimuthal angles
is required in order to complete the Bernoulli sampling. This bit can be sent down the binomial tree at the cost of $O(\log n)$ time
if we reverse the arrows in Figure~\ref{binomial} and reorder the transmissions from the
edge marked $\lceil \, \log_2 n \rceil$ (which is \mbox{$\lceil \, \log_2 8 \rceil=3$} in the figure)
down to the edges marked~$1$.

If we consider a nonstandard model in which we only care about what happens until all parties have
produced their output, we can modify the above protocol to require only one-way communication on each
of the links, namely up the binomial tree, with no increase (in~fact a small decrease)
in~expected communication
and time complexities before the final output has been produced.
For this, we simply remove the leader's notification to all other parties of whether or not
the simulation has been completed. This means that all parties will indefinitely continue to provide the leader
(who will pay no attention!)\ with increasingly precise approximations of the sum of their azimuthal angles,
but this useless activity will take place after all parties have produced their output. Indeed,
all parties other than
the leader can output their randomly selected $+1$ or $-1$ at the very beginning of the protocol,
and the leader can output his answer as soon as \emph{he} knows that the Bernoulli sampling
(Algorithm~\ref{algo-Bernoulli}) has been successful.

Of course, we could have parallelized line~\ref{step:Bern} of Algorithm~\ref{algo-final}
even in the case of non-equatorial
measurements. However, this would not have impacted significantly on the overall time complexity
of our general solution, which remains $O(n \log n)$.

\subsection{Only the leader needs to be probabilistic}\label{app:deterministic}
It is easy to modify almost all our protocols to require randomness only from the leader,
all other parties being purely deterministic.
For this, notice that the total expected amount of randomness is only $O(n)$,
which is negligible compared to the total number of bits that have to be communicated.
Hence, each time one party needs a random bit, he can ask the leader to provide~it.
This will only increase the communication cost by an expected $O(n)$ bits,
which has no effect on the overall asymptotic communication complexity of our protocols.
The same remark applies to the time required by our protocols in the parallel model,
with the exception of the case of equatorial measurements, in which an $O(\log^2\!n)$
expected time requires all parties other than the leader to choose their unbiased random output in parallel.

\section{Conclusion, discussion and open problems}\label{discussion}
We have addressed the problem of simulating the effect of arbitrary independent von Neumann measurements
on the qubits forming the general GHZ state \mbox{$\raisebox{0.04ex}[0pt][0pt]{\small$\textstyle \frac{1}{\sqrt{2}}$} | 0^n \rangle + \raisebox{0.04ex}[0pt][0pt]{\small$\textstyle \frac{1}{\sqrt{2}}$} | 1^n \rangle$} distributed
among $n$ parties. Rather than doing the actual quantum measurements, the parties must sample the
exact GHZ probability distribution by purely classical means, which necessarily requires communication in view
of Bell's theorem. Our main objective was to find a protocol that solves this conundrum with a finite amount
of expected communication, which had only been known previously to be possible when the von Neumann measurements
are restricted to being equatorial (a~severe limitation indeed).
Our solution needs only an expectation of $O(n^2)$ bits of
communication, which can be dispatched in $O(n \log n)$ expected time if bits can be sent in parallel
according to a realistic scenario in which
nobody has to send or receive more than one bit in any given step. We~also improved on the former art
in the case of equatorial measurements, with expectations of $O(n \log n)$ bits of communication and $O(\log^2\!n)$ parallel time.

Knuth and Yao~\cite{knuthyao76} initiated the study of the
complexity of generating random integers (or~bit strings) with a given probability distribution~$p$,
assuming only the availability of a source of unbiased
identically independently distributed random bits.
They showed that any sampling algorithm must use an expected number
of bits at least equal to the entropy $\sum_b p(b) \log_2 (1/p(b))$ of the distribution,
and that the best algorithm does not need more than two additional bits.
For further results on the bit model in random variate generation,
see Ref.~\cite[Chap.~XIV]{devbook86} and Ref.~\cite{DG15a}.

The GHZ distribution has an entropy no larger than~$n$,
and therefore Knuth and Yao have shown that it could be sampled with no more than \mbox{$n+2$} expected random bits if all
the parameters were concentrated in a single place.
Even though we have studied the problem of sampling this distribution
in a setting in which the defining parameters (here the description of the
von Neumann measurements) are distributed among $n$ parties,
and despite the fact that our main purpose was to minimize
communication between these parties,
we were able to succeed with \mbox{$6n+17$} expected random bits,
which is just above six times the bound of Knuth and Yao.
The~amount of randomness required by our protocols does not depend significantly
on the actual measurements they have to simulate, as discussed at the end
of Section~\ref{sct:analysis}. However, some sets of measurements
entail a probability distribution $p$ whose entropy $H(p)$ is much smaller than~$n$.
In~the extreme case of having all measurements in the computational basis,
$H(p)$ is a single bit! Can there be protocols that
succeed with as few as \mbox{$H(p)+2$} expected random bits,
thus meeting the bound of Knuth and Yao,
or failing this as few as $O(H(p))$ expected random bits,
no matter how small $H(p)$ is for the given set of von Neumann measurements?
Notice that all the protocols presented here require $\Omega(n)$ random bits
since they ask each party to sample independently at least once a Rademacher random variable,
a hurdle that can only be alleviated in the case of measurements in the computational basis.
It~\emph{may} be that this problem can be solved if we put the leader in charge
of drawing \emph{all} the Rademachers in a single batch.
But what would be the cost in terms of communication from the other parties,
who will need to send sufficiently precise approximations of their elevation angles~$\varphi_j$
to the leader, rather than the much easier task of generating their own Rademachers locally?

Are our protocols optimal in terms of the required amount of communication?
Could we simulate arbitrary von Neumann measurements as efficiently as in the case
of equatorial measurements, i.e.~with $O(n \log n)$ expected bits of communication?
We~leave this as open question, but point out that  Broadbent, Chouha and Tapp have proved
an $\Omega(n \log n)$ lower bound on the \emph{worst case} communication complexity of simulating
measurements on \mbox{$n$-partite} GHZ states~\cite{BCT09}, a result that holds even for equatorial measurements,
and even under the promise that \mbox{$\cos \sum_{j=1}^n \theta_j = \pm 1$}~\cite{MarcPC}.

As a recent development, which will be the subject of a follow-up paper~\cite{DG15b},
we have discovered how to handle \emph{arbitrary} \mbox{$n$-partite} states,
such as the tripartite $W$ state
\mbox{$\raisebox{0.04ex}[0pt][0pt]{\small$\textstyle \frac{1}{\sqrt{3}}$} | 100 \rangle + \raisebox{0.04ex}[0pt][0pt]{\small$\textstyle \frac{1}{\sqrt{3}}$} | 010 \rangle + \raisebox{0.04ex}[0pt][0pt]{\small$\textstyle \frac{1}{\sqrt{3}}$} | 001 \rangle$}
and its \mbox{$n$-partite} generalization
\[ W_n = {\textstyle \frac{1}{\sqrt{n}}} \, \big( | 10^{n-1} \rangle + | 010^{n-2} \rangle + | 0010^{n-3} \rangle +\cdots+  | 0^{n-1}1 \rangle \big) \, , \]
in which each of the $n$ parties is given one of the qubits.
Although the general simulation process is rather more complicated and slightly less efficient
than in the case of the GHZ state, the effect of $n$ independent von Neumann measurements
on any \mbox{$n$-partite} state (even a mixed state) distributed among $n$ participants can be
simulated with an expectation of \mbox{$(5+o(1))n^2$} bits of communication.
Only the leader needs access to a source of randomness and an expectation of \mbox{$(4+o(1))n^2$}
unbiased identically independently distributed bits suffices to carry out the simulation.

We leave for further research the problem of simulating
arbitrary \emph{positive-operator-valued measurements} (POVMs)
on the single-qubit shares of GHZ states (or~on more general multipartite states),
as well as the problem of simulating multipartite entanglement
(other than the already-solved equatorial von Neumann measurements on the tripartite GHZ state~\cite{BG11})
with \emph{worst-case} bounded classical communication.

\appendix

\begin{center}
\Large{\textbf{Appendices}}\vspace{-2mm}
\end{center}

\section{Convex decomposition of the GHZ distribution}\label{App-decomp}
Our simulation of the GHZ distribution hinges upon its decomposition into a convex combination of two
sub-distributions, which is stated as Equation~(\ref{eq:p}) at the beginning of Section~\ref{sampling},
\begin{displaymath}\textstyle
p(b)= \cos^2\!\big(\frac{\theta}{2}\big) \, p_1(b)+\sin^2\!\big(\frac{\theta}{2}\big) \, p_2(b) \, ,
\end{displaymath}
in which the coefficients $\cos^2(\theta/2)$ and $\sin^2(\theta/2)$
depend only on the equatorial part of the measurements,
whereas the sub-distributions depend only on their real part.
This decomposition was obtained by one of us~\cite{gra2011mem,gra2012}, albeit in the usual computer science language in which
von Neumann measurements are presented as a unitary transformation followed by a measurement
in the computational basis.
For completeness, here we derive this decomposition directly in the language of von Neumann measurements.

First, let us recall some facts, including some already mentioned in Section~\ref{sampling}.
We~begin with a $2\times 2$ von Neumann measurement, which can be written as
\begin{displaymath}
M = x\sigma_1 + y\sigma_2 + z\sigma_3 = x\left(
\begin{array}{cc}
0 & 1 \\
1 & 0 \\
\end{array} \right)
+y\left(
\begin{array}{cc}
0 & -\imath \\
\imath & 0 \\
\end{array} \right)
+z\left(
\begin{array}{cc}
1 & 0 \\
0 & -1 \\
\end{array} \right)=
\left(
\begin{array}{cc}
z & x-\imath y \\
x+\imath y & -z \\
\end{array} \right) \, ,
\end{displaymath}
where \mbox{$x^2+y^2+z^2=1$}.  Thus, using spherical coordinates
\mbox{$(\theta,\varphi)\in[0,2\pi)\times[-\pi\slash 2,\pi\slash 2]$}, the \mbox{parameters} \mbox{$(x,y,z)$} can be written as
\begin{eqnarray*}
x&=&\cos\theta\cos\varphi\\*
y&=&\sin\theta\cos\varphi\\*
z&=&\sin\varphi
\end{eqnarray*}
so that
\begin{displaymath}
M=\left(\begin{array}{cc}
\sin\varphi & e^{-\imath\theta} \cos\varphi\\
 e^{\imath\theta} \cos\varphi & -\sin\varphi
\end{array}\right) \, .
\end{displaymath}
The spectra (set of eigenvalues) of $M$ is $\{-1,+1\}$ and the unitary operator $U$ that diagonalizes $M$ is given by
\begin{displaymath}
U=\left(\begin{array}{cc}
\alpha & -\bar{\beta}\\
-\beta & -\bar{\alpha}
\end{array}\right)
\end{displaymath}
with
\begin{displaymath}
\alpha  =  \cos\Big(\frac{\varphi}{2}-\frac{\pi}{4}\Big)~~\text{and}~~\beta =e^{\imath\theta} \sin\Big(\frac{\varphi}{2}-\frac{\pi}{4}\Big) \, .
\end{displaymath}
In other words, we have
\begin{displaymath}
M=U\left(\begin{array}{cc}
1 & 0\\
0 & -1
\end{array}\right)U^{\dagger}.
\end{displaymath}
The density matrix representing the GHZ state can be decomposed as
\begin{displaymath}
\rho = \ket{\Psi_n}\bra{\Psi_n} = \frac{1}{2}\left(\bigotimes_{j=1}^{n}{\left(\begin{array}{cc}
1 & 0\\
0 & 0
\end{array}\right)}+\bigotimes_{j=1}^{n}{\left(\begin{array}{cc}
0 & 1\\
0 & 0
\end{array}\right)}+\bigotimes_{j=1}^{n}{\left(\begin{array}{cc}
0 & 0\\
1 & 0
\end{array}\right)}+\bigotimes_{j=1}^{n}{\left(\begin{array}{cc}
0 & 0\\
0 & 1
\end{array}\right)}\right) \, .
\end{displaymath}

Before analysing the joint probability function, we point out that
\begin{displaymath}
\ket{b}\bra{b} = \left(\begin{array}{cc}
\delta_{+1}(b) & 0\\
0 & \delta_{-1}(b)
\end{array}\right) ,
\end{displaymath}
where
\mbox{$b \in \{-1,+1\}$},
\mbox{$\ket{+1}= {\,1\, \choose \,0\,}$},
\mbox{$\ket{-1}={{\,0\,} \choose {\,1\,}}$}
and
$\delta$ is the Kronecker delta function
\mbox{$\delta_{x}(y)=1$} if~\mbox{$x=y$} and \mbox{$\delta_{x}(y)=0$} if \mbox{$x\neq y$}.
We~also invite the reader to verify that
\begin{eqnarray*}
\mathrm{Tr}\Bigg(\bigg(\ket{b}\bra{b}\bigg)U\left(\begin{array}{cc}
1 & 0\\
0 & 0
\end{array}\right)U^{\dagger}\Bigg) & = & |\alpha|^2\delta_{+1}(b)+|\beta|^2\delta_{-1}(b)\\
&=&\cos^2\!\bigg(\frac{1}{2}\Big(\varphi-\frac{\pi}{2}b\Big)\bigg)\\[2ex]
\mathrm{Tr}\Bigg(\bigg(\ket{b}\bra{b}\bigg)U\left(\begin{array}{cc}
0 & 1\\
0 & 0
\end{array}\right)U^{\dagger}\Bigg) & = & -\alpha\beta\delta_{+1}(b)+\alpha\beta\delta_{-1}(b)\\
&=&-e^{\imath\theta}\sin\!\bigg(\frac{1}{2}\Big(\varphi-\frac{\pi}{2}b\Big)\bigg)\cos\!\bigg(\frac{1}{2}\Big(\varphi-\frac{\pi}{2}b\Big)\bigg)\\[2ex]
\mathrm{Tr}\Bigg(\bigg(\ket{b}\bra{b}\bigg)U\left(\begin{array}{cc}
0 & 0\\
1 & 0
\end{array}\right)U^{\dagger}\Bigg) & = & -\bar{\alpha}\bar{\beta}\delta_{+1}(b)+\bar{\alpha}\bar{\beta}\delta_{-1}(b)\\
&=&-e^{-\imath\theta}\sin\!\bigg(\frac{1}{2}\Big(\varphi-\frac{\pi}{2}b\Big)\bigg)\cos\!\bigg(\frac{1}{2}\Big(\varphi-\frac{\pi}{2}b\Big)\bigg)\\[2ex]
\mathrm{Tr}\Bigg(\bigg(\ket{b}\bra{b}\bigg)U\left(\begin{array}{cc}
0 & 0\\
0 & 1
\end{array}\right)U^{\dagger}\Bigg) & = & |\beta|^2\delta_{+1}(b)+|\alpha|^2\delta_{-1}(b)\\
&=&\sin^2\!\bigg(\frac{1}{2}\Big(\varphi-\frac{\pi}{2}b\Big)\bigg) \, .
\end{eqnarray*}
For convenience, let
\begin{displaymath}
E_1=\left(\begin{array}{cc}
1 & 0\\
0 & 0
\end{array}\right),~E_2=\left(\begin{array}{cc}
0 & 1\\
0 & 0
\end{array}\right),~E_3=\left(\begin{array}{cc}
0 & 0\\
1 & 0
\end{array}\right)~\text{and}~E_4=\left(\begin{array}{cc}
0 & 0\\
0 & 1
\end{array}\right).
\end{displaymath}
Given $n$ von Neumann measurements $M_j$, \mbox{$1 \le j \le n$},
the joint probability $p(b)$ of obtaining \mbox{$b\in \{-1,+1\}^n$} as a result of applying these measurements
on an \mbox{$n$-par}\-tite GHZ state is
\begin{eqnarray*}
p(b)&=&|\bra{b}U\ket{\Psi_n}|^2\\
&=&\mathrm{Tr}\Big(\bra{b}U\ket{\Psi_n}\big(\bra{b}U\ket{\Psi_n}\big)^{\dagger}\Big)\\
&=&\mathrm{Tr}\Big(\bra{b}U\ket{\Psi_n}\bra{\Psi_n}U^{\dagger}\ket{b}\Big)\\
&=&\mathrm{Tr}\bigg(\Big(\ket{b}\bra{b}\Big)U\rho{}\,U^{\dagger}\bigg)\\
&=&\frac{1}{2}\mathrm{Tr}\Bigg(\bigotimes_{j=1}^{n}{\ket{b_j}\bra{b_j}}\Bigg(\bigotimes_{j=1}^{n}{U_j E_1U_j^{\dagger}}+\bigotimes_{j=1}^{n}{U_j E_2 U_j^{\dagger}}+\bigotimes_{j=1}^{n}{U_j E_3 U_j^{\dagger}}+\bigotimes_{j=1}^{n}{U_j E_4 U_j^{\dagger}}\Bigg)\Bigg)\\
&=&\frac{1}{2}\sum_{i=1}^4{\prod_{j=1}^{n}{\mathrm{Tr}\Bigg(\left(\begin{array}{cc}
\delta_{+1}(b_j) & 0\\
0 & \delta_{-1}(b_j)
\end{array}\right)U_j E_i U_j^{\dagger}\Bigg)}} \,.
\end{eqnarray*}
Putting these equations together, we have:
\begin{eqnarray*}
\prod_{j=1}^{n}{\mathrm{Tr}\Bigg(\left(\begin{array}{cc}
\delta_{+1}(b_j) & 0\\
0 & \delta_{-1}(b_j)
\end{array}\right)U_j E_1 U_j^{\dagger}\Bigg)}&=&\prod_{j=1}^{n}{\cos^2\!\bigg(\frac{1}{2}\Big(\varphi_j-\frac{\pi}{2}b_j\Big)\bigg)}\label{pp1} ~~\isdef~~ f_1\\
\prod_{j=1}^{n}{\mathrm{Tr}\Bigg(\left(\begin{array}{cc}
\delta_{+1}(b_j) & 0\\
0 & \delta_{-1}(b_j)
\end{array}\right)U_j E_2 U_j^{\dagger}\Bigg)}&=&\prod_{j=1}^{n}{-e^{\imath\theta_j}\sin\!\bigg(\frac{1}{2}\Big(\varphi_j-\frac{\pi}{2}b_j\Big)\bigg)\cos\!\bigg(\frac{1}{2}\Big(\varphi_j-\frac{\pi}{2}b_j\Big)\bigg)}\label{pp2} ~~\isdef~~ f_2\\
\prod_{j=1}^{n}{\mathrm{Tr}\Bigg(\left(\begin{array}{cc}
\delta_{+1}(b_j) & 0\\
0 & \delta_{-1}(b_j)
\end{array}\right)U_j E_3 U_j^{\dagger}\Bigg)}&=&\prod_{j=1}^{n}{-e^{-\imath\theta_j}\sin\!\bigg(\frac{1}{2}\Big(\varphi_j-\frac{\pi}{2}b_j\Big)\bigg)\cos\!\bigg(\frac{1}{2}\Big(\varphi_j-\frac{\pi}{2}b_j\Big)\bigg)}\label{pp3}~~\isdef~~ f_3\\
\prod_{j=1}^{n}{\mathrm{Tr}\Bigg(\left(\begin{array}{cc}
\delta_{+1}(b_j) & 0\\
0 & \delta_{-1}(b_j)
\end{array}\right)U_j E_4 U_j^{\dagger}\Bigg)}&=&\prod_{j=1}^{n}{\sin^2\!\bigg(\frac{1}{2}\Big(\varphi_j-\frac{\pi}{2}b_j\Big)\bigg)}\label{pp4}~~\isdef~~ f_4\,.
\end{eqnarray*}
Thus, $p(b)=\frac{1}{2}(f_1+f_2+f_3+f_4)$.

Keeping in mind that
\begin{displaymath}
f_2+f_3 = 2\cos\!\bigg(\sum_{j=1}^{n}{\theta_j}\bigg)\prod_{j=1}^{n}{-\sin\!\bigg(\frac{1}{2}\Big(\varphi_j-\frac{\pi}{2}b_j\Big)\bigg)\cos\!\bigg(\frac{1}{2}\Big(\varphi_j-\frac{\pi}{2}b_j\Big)\bigg)}
\end{displaymath}
and
\begin{displaymath}
x^2+y^2+2xy\cos \gamma = (x+y)^2\cos^2(\gamma/2)+(x-y)^2\sin^2(\gamma/2)
\end{displaymath}
for all real numbers $x$ and $y$, and angle $\gamma$,
it follows that
\begin{eqnarray*}
p(b)&=&\cos^2\!\Bigg(\frac{1}{2}\sum_{j=1}^{n}{\theta_j}\Bigg)\Bigg(\frac{\prod_{j=1}^{n}{\cos\big(\frac{1}{2}\big(\varphi_j-\frac{\pi}{2}b_j\big)\big)}+\prod_{j=1}^{n}{-\sin\big(\frac{1}{2}\big(\varphi_j-\frac{\pi}{2}b_j\big)\big)}}{\sqrt{2}}\Bigg)^{\!\!2}\,+\\*
&&\sin^2\!\Bigg(\frac{1}{2}\sum_{j=1}^{n}{\theta_j}\Bigg)\Bigg(\frac{\prod_{j=1}^{n}{\cos\big(\frac{1}{2}\big(\varphi_j-\frac{\pi}{2}b_j\big)\big)}-\prod_{j=1}^{n}{-\sin\big(\frac{1}{2}\big(\varphi_j-\frac{\pi}{2}b_j\big)\big)}}{\sqrt{2}}\Bigg)^{\!\!2}  \, .
\end{eqnarray*}

If we define
\begin{align*}
\mathrm{a}_1(b)&=\prod_{j=1}^{n} \textstyle {\cos\big(\frac{1}{2}\big(\varphi_j-\frac{\pi}{2}b_j\big)\big)},&
\mathrm{a}_2(b)&=\prod_{j=1}^{n} \textstyle {-\sin\big(\frac{1}{2}\big(\varphi_j-\frac{\pi}{2}b_j\big)\big)} \, , \\
p_1(b)&=\frac{1}{2}\big(\mathrm{a}_1(b)+\mathrm{a}_2(b)\big)^2,&p_2(b)&=\frac{1}{2}\big(\mathrm{a}_1(b)-\mathrm{a}_2(b)\big)^2 \, ,
\end{align*}
this is precisely the convex decomposition
\begin{displaymath}\textstyle
p(b)=
\cos^2\!\big(\frac{\theta}{2}\big) \, p_1(b)+\sin^2\!\big(\frac{\theta}{2}\big) \, p_2(b) \, ,
\end{displaymath}
where \mbox{$\theta = \sum_{j=1}^{n} \theta_j$},
that was given as Equation~(\ref{eq:p}) at the beginning of Section~\ref{sampling}.

As a ``reality check'', we analyse this formula for the special case of equatorial measurements,
in which all elevation angles vanish.
The formulas for $a_1(b)$ and $a_2(b)$, and therefore those for $p_1(b)$ and $p_2(b)$,
become very simple when $\varphi_j = 0$ for all~$j$.
For any \mbox{$b=(b_1,\ldots,b_n) \in \{-1,+1\}^n$}, let us define
\smash{\mbox{$s(b) =  \prod_{j=1}^n b_j \in \{-1,+1\}$}}
and
\mbox{$X = \{b \in \{-1,+1\}^n \mid s(b) = +1 \}$}.
It~is easy to see that
\mbox{$a_1(b) = 2^{-n/2}$} and \mbox{$a_2(n) = (-1)^{s(b)} 2^{-n/2}$}.
Therefore,
\[
p_1(b) = \left\{ \begin{array}{ll} 2^{1-n} & \mbox{if~} b \in X \\ 0 & \mbox{if~} b \not\in X \end{array} \right.
\mbox{~~~and~~~}
p_2(b) = \left\{ \begin{array}{ll} 0 & \mbox{if~} b \in X \\ 2^{1-n} & \mbox{if~} b \not\in X \,. \end{array} \right.
\]
Hence,
\begin{equation}\label{eq:eqp}
p(b) = \left\{ \begin{array}{ll} 2^{1-n} \cos^2\!\big(\frac{\theta}{2}\big) & \mbox{if~} b \in X \\[1.5ex]
             2^{1-n} \sin^2\!\big(\frac{\theta}{2}\big) & \mbox{if~} b \not\in X \,. \end{array} \right.
\end{equation}
Thus, in the case of equatorial measurements, we obtain a uniformly distributed \mbox{$b \in X$} with probability
$\cos^2(\theta/2)$ or a uniformly distributed \mbox{$b \in \{-1,+1\}^n \setminus X$}
with complementary probability $\sin^2(\theta/2)$.
From this, it follows immediately that the expected value of the product of the $b_j$'s is equal to
the cosine of the sum of the azimuthal angles because
\[ \mathbf{E} \Big\{ \prod_{j=1}^n b_j  \Big\} =
\textstyle \cos^2\!\big(\frac{\theta}{2}\big) \times (+1) +  \sin^2\!\big(\frac{\theta}{2}\big) \times (-1) = \cos^2\!\big(\frac{\theta}{2}\big) - \sin^2\!\big(\frac{\theta}{2}\big) = \cos \theta = \displaystyle \cos\!\Big(\sum_{j=1}^{n} \theta_j\Big)\, . \]
It~follows equally easily that \mbox{$\mathbf{E} \big\{ \prod_{j \in J} b_j \big\} = 0$} for any nonempty
\mbox{$J \subsetneq \{1,\ldots,n\}$}, and therefore all the marginal probability
distributions obtained by tracing out one or more of the parties are uniform.
Those well-known facts were indeed the formulas used in the prior art of simulating equatorial measurements
on GHZ states~\cite{BBG10,BG11,BK12}.

\section{Approximations and truncations of products}\label{App-error}
In this appendix, we restrict our attention to the multiplication of real numbers in the interval~\mbox{$[0,1]$}
because this is what is relevant to the analysis of the parallel model, in which we need to approximate the
product of sines and cosines sent up the binomial tree of Figure~\ref{binomial}.
It~is sufficient, again for simplicity, to concentrate on positive numbers because the signs
can be percolated independently up the binomial tree.

Consider any \mbox{$t \in [0,1]$} and
positive integer~$\ell$. Recall from Definition~\ref{def:approxtrunc} that the \mbox{$\ell$-bit}
truncation of $t$ is $\lfloor t2^{\ell} \rfloor/2^{\ell}$ because $t$ is nonnegative.
This \mbox{$\ell$-bit} truncation is obviously an \mbox{$\ell$-bit} approximation as well:
\begin{displaymath}
\bigg|\frac{\lfloor t2^{\ell} \rfloor}{2^{\ell}}-t\bigg|\leq \frac{1}{2^{\ell}}.
\end{displaymath}

Suppose now that we have two numbers $x_j$ and $y_j$ at level $j$ in the binomial tree inherent to Algorithm~\ref{algo-parallel}, such that both numbers lie in interval~\mbox{$[0,1]$}.
We~can express $x_j$ and $y_j$ recursively using the numbers $x_{j-1,1}$, $x_{j-1,2}$, $y_{j-1,1}$, and $y_{j-1,2}$ as follows:
\begin{eqnarray*}
x_j&=&\bigg\lfloor \frac{\lfloor x_{j-1,1}2^{\ell} \rfloor}{2^{\ell}}\frac{\lfloor x_{j-1,2}2^{\ell} \rfloor}{2^{\ell}}2^{\ell}\bigg \rfloor\frac{1}{2^{\ell}}\\
y_j&=&\bigg\lfloor \frac{\lfloor y_{j-1,1}2^{\ell} \rfloor}{2^{\ell}}\frac{\lfloor y_{j-1,2}2^{\ell} \rfloor}{2^{\ell}}2^{\ell}\bigg \rfloor\frac{1}{2^{\ell}} \, . \\
\end{eqnarray*}
We use $\epsilon_j$ for the error at level $j$ on the product $x_jy_j$; in other words,
\begin{displaymath}
\bigg|\bigg\lfloor \frac{\lfloor x_j 2^{\ell} \rfloor}{2^{\ell}}\frac{\lfloor y_j 2^{\ell} \rfloor}{2^{\ell}}2^{\ell}\bigg \rfloor\frac{1}{2^{\ell}}-x_j y_j \bigg|= \epsilon_j \Longleftrightarrow\bigg|\bigg\lfloor \frac{\lfloor x_j 2^{\ell} \rfloor}{2^{\ell}}\frac{\lfloor y_j 2^{\ell} \rfloor}{2^{\ell}}2^{\ell}\bigg \rfloor\frac{1}{2^{\ell}}-\frac{\lfloor x_{j-1}2^{\ell} \rfloor}{2^{\ell}}\frac{\lfloor y_{j-1}2^{\ell} \rfloor}{2^{\ell}}\bigg|= \epsilon_j
\end{displaymath}
Before bounding $\epsilon_j$ from above, we notice that,
\begin{eqnarray*}
\bigg|x_j-\frac{\lfloor x_j 2^{\ell} \rfloor}{2^{\ell}}\bigg|&=&\bigg|\bigg\lfloor \frac{\lfloor x_{j-1,1}2^{\ell} \rfloor}{2^{\ell}}\frac{\lfloor x_{j-1,2}2^{\ell} \rfloor}{2^{\ell}}2^{\ell}\bigg \rfloor\frac{1}{2^{\ell}}-\frac{\lfloor x_{j-1,1}2^{\ell} \rfloor}{2^{\ell}}\frac{\lfloor x_{j-1,2}2^{\ell} \rfloor}{2^{\ell}}\bigg|\\[2ex]
&=&\epsilon_{j-1}
\end{eqnarray*}
and the same for $y_j$.  We~also establish the following inequality
\begin{eqnarray*}
\bigg|\frac{\lfloor x_j 2^{\ell} \rfloor}{2^{\ell}}\frac{\lfloor y_j 2^{\ell} \rfloor}{2^{\ell}}-x_j y_j \bigg|&=&\bigg|\frac{\lfloor x_j 2^{\ell} \rfloor}{2^{\ell}}\frac{\lfloor y_j 2^{\ell} \rfloor}{2^{\ell}}-x_j \frac{\lfloor y_j 2^{\ell} \rfloor}{2^{\ell}}+x_j\frac{\lfloor y_j 2^{\ell} \rfloor}{2^{\ell}}-x_jy_j \bigg|\\
&\leq&\bigg|\frac{\lfloor x_j 2^{\ell} \rfloor}{2^{\ell}}\frac{\lfloor y_j 2^{\ell} \rfloor}{2^{\ell}}-x_j\frac{\lfloor y_j 2^{\ell} \rfloor}{2^{\ell}}\bigg|+\bigg|x_j\frac{\lfloor y_j 2^{\ell} \rfloor}{2^{\ell}}-x_j y_j \bigg|\\
&=&\bigg|\frac{\lfloor y_j 2^{\ell} \rfloor}{2^{\ell}}\bigg|\bigg|\frac{\lfloor x_j 2^{\ell} \rfloor}{2^{\ell}}-x_j\bigg|+|x_j|\bigg|\frac{\lfloor y_j 2^{\ell} \rfloor}{2^{\ell}}-y_j\bigg|\\
&=&\epsilon_{j-1}+\epsilon_{j-1}\\
&=&2\epsilon_{j-1} \,.
\end{eqnarray*}
Now we have that
\begin{eqnarray*}
\epsilon_j&=&\bigg|\bigg\lfloor \frac{\lfloor x_j 2^{\ell} \rfloor}{2^{\ell}}\frac{\lfloor y_j 2^{\ell} \rfloor}{2^{\ell}}2^{\ell}\bigg \rfloor\frac{1}{2^{\ell}}-x_jy_j \bigg|\\
&=&\bigg|\bigg\lfloor \frac{\lfloor x_j 2^{\ell} \rfloor}{2^{\ell}}\frac{\lfloor y_j 2^{\ell} \rfloor}{2^{\ell}}2^{\ell}\bigg \rfloor\frac{1}{2^{\ell}}-\frac{\lfloor x_j 2^{\ell} \rfloor}{2^{\ell}}\frac{\lfloor y_j 2^{\ell} \rfloor}{2^{\ell}}+\frac{\lfloor x_j 2^{\ell} \rfloor}{2^{\ell}}\frac{\lfloor y_j 2^{\ell} \rfloor}{2^{\ell}}-x_jy_j \bigg|\\
&\leq&\bigg|\bigg\lfloor \frac{\lfloor x_j 2^{\ell} \rfloor}{2^{\ell}}\frac{\lfloor y_j 2^{\ell} \rfloor}{2^{\ell}}2^{\ell}\bigg \rfloor\frac{1}{2^{\ell}}-\frac{\lfloor x_j 2^{\ell} \rfloor}{2^{\ell}}\frac{\lfloor y_j 2^{\ell} \rfloor}{2^{\ell}}\bigg|+\bigg|\frac{\lfloor x_j 2^{\ell} \rfloor}{2^{\ell}}\frac{\lfloor y_j 2^{\ell} \rfloor}{2^{\ell}}-x_j y_j \bigg|\\
&\leq&\frac{1}{2^{\ell}}+2\epsilon_{j-1}.
\end{eqnarray*}
If there are $m=\lceil\,\log_2 n\rceil$ levels,
\begin{displaymath}
\epsilon_{m}\leq \frac{1}{2^{\ell}}\sum_{j=0}^{m}{2^j} < \frac{2^{\lceil\,\log_2 n\rceil}}{2^{\ell-1}} \, .
\end{displaymath}
It follows that a \mbox{$k$-bit} approximation of the product of $n$ real numbers in interval \mbox{$[0,1]$}, which corresponds to $\epsilon_{m} \le 2^{-k}$,
is obtained if we truncate each intermediate subproduct to \mbox{$\ell = k+1+\lceil\,\log_2 n\rceil$} bits.

\section*{Acknowledgements}
We wish to thank Marc Kaplan and Nicolas Gisin for stimulating discussions about the simulation of entanglement.
Furthermore, Marc has carefully read Ref.~\cite{gra2011mem}, in which the decomposition of the GHZ distribution
as a convex combination of two sub-distributions was first accomplished,
and he has pointed out that the lower bound from Ref.~\cite{BCT09} applies
even in the case of equatorial measurements.
Alain Tapp pointed out that the entropy of the GHZ distribution can be as small as one bit in the case of
measurements in the computational basis.

G.\,B.~is supported in part by the Natural Sciences
and Engineering Research Council of Canada (\textsc{Nserc}),
the Canada Research Chair program,
the Canadian Institute for \mbox{Advanced} Research (\textsc{Cifar}),
the Institut transdisciplinaire d'information quantique (\textsc{Intriq}).
Part of this research was accomplished while G.\,B.~was a Fellow at the
Institute for Theoretical Studies (ITS) of ETH~Z\"urich.
L.\,D.~is~supported in part by \textsc{Nserc}
and Fonds de recherche du Qu\'ebec -- Nature et techno\-logies (\textsc{Frqnt}).


\begin{thebibliography}{10}

\bibitem{BBG10}
J.-D. Bancal, C. Branciard and N. Gisin,
``Simulation of equatorial von Neumann measurements on GHZ states
  using nonlocal resources'',
\emph{Advances in Mathematical Physics}
\textbf{2010}:293245,
2010.

\bibitem{bell64}
J.\,S. Bell,
``On~the Einstein-Podolsky-Rosen paradox'',
\emph{Physics}
\textbf{1}:195--200, 1964.

\bibitem{BG11}
C. Branciard and N. Gisin,
``Quantifying the nonlocality of Greenberger-Horne-Zeilinger
  quantum correlations by a bounded communication simulation protocol'',
\emph{Physical Review Letters}
\textbf{107}:020401, 2011.

\bibitem{QCC03}
G. Brassard,
``Quantum communication complexity'',
\emph{Foundations of Physics}
\textbf{33}(11):1593--1616, 2003.

\bibitem{bct99}
G. Brassard, R. Cleve and A. Tapp,
``Cost of exactly simulating quantum entanglement with classical
  communication'',
\emph{Physical Review Letters}
\textbf{83}:1874--1877, 1999.

\bibitem{DG15b}
G. Brassard, L. Devroye and C. Gravel,
``Simulation of entanglement and distributed sampling of quantum
  probability distributions'',
in preparation,
2015.

\bibitem{BK12}
G. Brassard and M. Kaplan,
``Simulating equatorial measurements on GHZ states with finite
  expected communication cost'',
\emph{Proceedings of 7th Conference on Theory of Quantum
  Computation, Communication, and Cryptography (TQC)}, Tokyo, pages 65--73, 2012.

\bibitem{BCT09}
A. Broadbent, P.\,R. Chouha and A. Tapp,
``The GHZ state in secret sharing and entanglement simulation'',
\emph{Proceedings of Third International Conference on Quantum,
  Nano and Micro Technologies}, Canc\'un, pp.~59--62, 2009.

\bibitem{CGM00}
N. Cerf, N. Gisin and S. Massar,
``Classical teleportation of a quantum bit'',
\emph{Physical Review Letters}
\textbf{84}(11):2521--2524, 2000.

\bibitem{devbook86}
L. Devroye,
\emph{Non-Uniform Random Variate Generation}'',
Springer, New York, 1986.

\bibitem{DG15a}
L. Devroye and C. Gravel,
``Sampling with arbitrary precision'',
\texttt{http://arxiv.org/abs/}\linebreak
\texttt{1502.02539},
2015.

\bibitem{EPR35}
A. Einstein, B. Podolsky and N. Rosen,
``Can quantum-mechanical description of physical reality be considered
  complete?'',
\emph{Physical Review}
\textbf{47}:777--780, 1935.

\bibitem{gp10}
N. Gisin,
personal communication,
2010.

\bibitem{gra2011mem}
C. Gravel,
\emph{Structure de la distribution de probabilit\'e de l'\'etat GHZ sous
  l'action de mesures de von Neumann locales},
Master's thesis, Universit\'e de Montr\'eal,
\texttt{https://papyrus.bib.}\linebreak
\texttt{umontreal.ca/jspui/handle/1866/5511},
2011.

\bibitem{gra2012}
C. Gravel,
``Structure of the probability distribution for the GHZ quantum state
  under local von Neumann measurements'',
\emph{Quantum Physics Letters}
\textbf{1}(3):87--96, 2012.

\bibitem{GHZ89}
D.\,M. Greenberger, M.\,A. Horne and A. Zeilinger,
``Going beyond {B}ell's theorem'',
in~\emph{Bell's Theorem, Quantum Theory and  Conceptions of the Universe}
(M.~Kafatos, ed.), Kluwer Academic, Dordrecht,
pp.~69--72, 1989.

\bibitem{MarcPC}
M. Kaplan,
personal communication,
2013.

\bibitem{knuthyao76}
D.\,E. Knuth and A.\,C.-C. Yao,
``The complexity of nonuniform random number generation'',
in~\emph{Algorithms and Complexity: New Directions and Recent Results}
(J.\,F. Traub, ed.), Academic Press, New York,
pp.~357--428, 1976.
Reprinted in D.\,E. Knuth, \textit{Selected Papers on Analysis of
  Algorithms}, Cambridge University Press, 2000.

\bibitem{MBCC01}
S. Massar, D. Bacon, N. Cerf and R. Cleve,
``Classical simulation of quantum entanglement without local hidden
  variables'',
\emph{Physical Review~A}
\textbf{63}(5):052305, 2001.

\bibitem{maudlin92}
T. Maudlin,
``Bell's inequality, information transmission, and prism models'',
\emph{PSA:~Proceedings of the Biennial Meeting of the
  Philosophy of Science Association}, Chicago, pp.~404--417, 1992.

\bibitem{vN51}
J. von Neumann,
``Various techniques used in connection with random digits. Monte
  Carlo methods'',
\emph{National Bureau of Standards}
\textbf{12}:36--38, 1951.

\bibitem{rt09}
O. Regev and B. Toner,
``Simulating quantum correlations with finite communication'',
\emph{SIAM Journal on Computing}
\textbf{39}(4):1562--1580, 2009.

\bibitem{steiner00}
M. Steiner,
``Towards quantifying non-local information transfer: Finite-bit
  non-locality'',
\emph{Physics Letters~A}
\textbf{270}:239--244, 2000.

\bibitem{tb03}
B. Toner and D. Bacon,
``Communication cost of simulating Bell correlations'',
\emph{Physical Review Letters}
\textbf{91}:187904, 2003.

\end{thebibliography}
\end{document}